\documentclass[%
 aip,
 jmp,%
 amsmath,amssymb,
reprint,%
]{revtex4-1}

\usepackage{graphicx}% Include figure files
\usepackage{dcolumn}% Align table columns on decimal point
\usepackage{bm}% bold math
\usepackage{hyperref}
\usepackage{natbib}
\usepackage{epstopdf}

\begin{document}

%\markboth{Anish Rai, Ajit Mahata, Md. Nurujjaman \& Om Prakash}
%{Statistical properties of the aftershocks of stock market crashes}

%%%%%%%%%%%%%%%%%%%%% Publisher's Area please ignore %%%%%%%%%%%%%%%
%\catchline{}{}{}{}{}
%%%%%%%%%%%%%%%%%%%%%%%%%%%%%%%%%%%%%%%%%%%%%%%%%%%%%%%%%%%%%%%%%%%%

\title{Statistical properties of the aftershocks of stock market crashes revisited: Analysis based on the 1987 crash,  financial-crisis-2008 and COVID-19 pandemic}

\author{Anish Rai, Ajit Mahata and Md.Nurujjaman*}
\address{Department of Physics, National Institute of Technology Sikkim\\
Sikkim, 737139, India \\
 Email:anishrai412@gmail.com\\
 Email:ajitnonlinear@gmail.com\\
*Email: md.nurujjaman@nitsikkim.ac.in}

%\author{Ajit Mahata}

%\address{Department of Physics, National Institute of Technology Sikkim\\
%Sikkim, 737139, India\\
%ajitnonlinear@gmail.com}

%\author{Md.Nurujjaman}
%\address{Department of Physics, National Institute of Technology Sikkim\\
%Sikkim, 737139, India\\
%jaman\_nonlinear@yahoo.co.in}

\author{Om Prakash}
\address{Department of Mathematics, National Institute of Technology Sikkim\\
Sikkim, 737139, India\\
Email:om.prakash@nitsikkim.ac.in}

%\maketitle

%\begin{history}
%\received{Day Month Year}
%\revised{Day Month Year}
%\end{history}

\begin{abstract}
During any unique crisis, panic sell-off leads to a massive stock market crash that may continue for more than a day, termed as mainshock. The effect of a mainshock in the form of aftershocks can be felt throughout the recovery phase of stock price.  As the market remains in stress during recovery, any small perturbation leads to a relatively smaller aftershock. The duration of the recovery phase has been estimated using structural break analysis. We have carried out statistical analyses of 1987 stock market crash, 2008 financial crisis and 2020 COVID-19 pandemic considering the actual crash-times of the mainshock and aftershocks. Earlier, such analyses were done considering absolute one-day return, which cannot capture a crash properly. The results show that the mainshock and aftershock in the stock market follows the Gutenberg-Richter (GR) power law. Further, we obtained higher $\beta$ value for the COVID-19 crash compared to the financial-crisis-2008 from the GR law. This implies that the recovery of stock price during COVID-19 may be faster than the financial-crisis-2008. The result is consistent with the present recovery of the market from the COVID-19 pandemic. The analysis shows that the high magnitude aftershocks are rare, and low magnitude aftershocks are frequent during the recovery phase. The analysis also shows that the distribution $P(\tau_i)$ follows the generalized pareto distribution, i.e., $\displaystyle~P(\tau_i)\propto\frac{1}{\{1+\lambda(q-1)\tau_i\}^{\frac{1}{(q-1)}}}$, where $\lambda$ and $q$ are constants and $\tau_i$ is the inter-occurrence time. This analysis may help investors to restructure their portfolio during a market crash. 

%\keywords{Market crash; Gutenberg-Richter power law; Structural break; Financial crisis-2008; The COVID-19 pandemic}
\end{abstract}

%\ccode{PACS Nos.:05.45.Tp, 89.65.Gh}
\maketitle
\section{Introduction}
\label{sec:intro}
Complex dynamical systems, such as stock market, earth crust, climate, ocean show rare phenomena in the form of price crash, earthquake, storm and tsunami under certain crisis~\cite{sornette2017stock,mcnamara2015earthquake,sarlis2018natural,pisarenko2003characterization,nguyen2017vietnam,piccoli2017stock,zhang2009estimating,sornette1998hierarchical,sornette1996stock,mahata2021characteristics}. Some of these rare events which can be termed as a main-shock, leads to a series of smaller aftershocks~\cite{sornette2017stock,sornette1996stock,abe2004aging,siokis2012stock}. The dynamics of such systems can be characterized statistically by power-law distributions~\cite{bak2002unified,bai2010power,gabaix2007unified,siokis2012stock,selccuk2004financial,kapopoulos2005stock}. In the stock market, a rare event occurs in the form of a major crash due to the emergence of a new and novel risk factor, for which investors are not prepared~\cite{albuquerque2020resiliency,mazur2020covid}. The new unique risk leads to a wide fluctuation in price and volume because of the investors' irrational behavior, and causes a massive crash~\cite{mazur2020covid,lyocsa2020fear,baker2020unprecedented,wang2009stock,barlevy2003rational}.

The reasons for the crash in stock market is the formation of instability due to herding behavior, speculation,  overvaluation and emergence of novel risk factors~\cite{chiang2010empirical,tan2008herding,white1990stock,choudhry1996stock,carlson2007brief,grant1990stock,mahata2021characteristics,mazur2020covid,lyocsa2020fear,baker2020unprecedented,wang2009stock,barlevy2003rational}. The financial crisis and pandemic have a devastating effect on all the economic sectors that lasts for several months to years. As a result stock market remains in stress till the economy starts to recover~\cite{sornette1997large,mazur2020covid,mahata2021modeling}. During the stressed period after the major crash, a number of aftershocks of comparatively smaller magnitudes occur due to investors' cautious approach in investment~\cite{siokis2012stock,selccuk2004financial,kapopoulos2005stock,netter1989stock,hong2003differences}. Understanding the properties of the mainshock and aftershocks are essential for the investment decision during a crisis. 

Statistical properties of the volatility of stock price, volume, daily returns of stock market have been extensively studied using the statistical tools like correlation, power law, multifractality and many other techniques~\cite{gopikrishnan2000statistical,oh2011statistical,liu1999statistical,mantegna1997stock,cont2001empirical,restocchi2019statistical,da2020relevant,noh2000model,hamao1990correlations,wang2012multifractal,turiel2003multifractal}. However, limited studies on the statistical properties of crash in stock price are carried out. ~\cite{siokis2012stock,selccuk2004financial,kapopoulos2005stock,potirakis2013dynamical,petersen2010market}. Further, these studies have some limitations with regard to the length of crashes and duration of its effect ~\cite{siokis2012stock,selccuk2004financial,kapopoulos2005stock}. Hence, a detailed statistical analysis is required to understand the stock market crash.

The analysis methods that are applied to study the geophysical phenomena have also been applied to understand the statistical properties of the price crash and subsequent aftershocks ~\cite{siokis2012stock,selccuk2004financial,kapopoulos2005stock}. The Gutenberg-Richter (GR) power law and Omori law are frequently used to explain such statistical properties ~\cite{siokis2012stock,selccuk2004financial,kapopoulos2005stock,potirakis2013dynamical,petersen2010market}. In these studies, the mainshock and aftershocks were identified for intra-day data with the one-minute absolute log return~\cite{petersen2010market}, and 60, 100 and 240 days data with the one-day absolute log return~\cite{selccuk2004financial,lillo2003power,siokis2012stock,kapopoulos2005stock,selccuk2004financial}. Omori law is also seen to hold not only for stock returns but also for stock volatilities~\cite{mu2008relaxation}. Models have also been developed to predict the crash in the stock market considering similarities between the stock return after a major crash with aftershock activities during an earthquake~\cite{gresnigt2015interpreting}. However, all these analyses have the following limitations:
\begin{enumerate}
\item As the market crash during a severe crisis can continue for more than a day~\cite{sornette1997large,sornette1996stock,sornette1998hierarchical}, the mainshock and aftershock cannot be limited to a single day or minute return.
\item	After the implementation of the circuit breakers in the market, it takes more than a day to complete an intense crash. Hence, taking one day fall as a mainshock or aftershock may be erroneous.  
\item	When one-day absolute log return is considered, both the crash and recovery act as a crash. 
\item	The duration of the aftershock sequence has been taken arbitrarily though the effect of major crash may stay for a longer/shorter period in the market and economy. Hence, identifying the proper duration of the mainshock is very essential to analyze the statistics of shocks.
\end{enumerate}

This paper aims to study the statistical properties of stock market crash by addressing the above limitations. We have estimated the actual crash-time of mainshock and aftershocks as the cumulative consecutive fall of a stock price. During the estimation of continuous fall, we have ignored the smaller intermediate weak recoveries. Maximum magnitude of these weak recoveries is 7.0\% of the mainshock. If the secondary crashes are larger than 7.0\% of the main crash, they are considered as an
aftershock. Further, the structural break analysis method~\cite{bai1998estimating,bai2003computation} is applied to estimate the duration of influence of the mainshock. The aftershocks that happened during this duration are considered for the analysis. We obtained that the aftershock sequence follows the GR power law. Adjusted $R^2$ and P-value from K-S test are calculated to verify that the data fits well with the GR power law. $R^2$ and P-value indicates how well the data fits with the power law. The P-value greater than 0.1 and 0.05 are considered as a good fit between the power law distribution and the data~\cite{clauset2009power,taleghani2019analysis}. We have calculated the values of $R^2$ and P-value to show that the power law distribution is a good fit to the data. We also intend to extract some new information about the possible occurrence of aftershocks during the latest stock market crash in 2020 due to the COVID-19 pandemic.

 The paper is organised as follows: Sec.~\ref{sec:meth} discusses about the methods used for the analysis. Sec.~\ref{sec:da} provides the data description. Sec.~\ref{sec:res} discusses about the results obtained from the analysis. Finally, Sec.~\ref{sec:con} includes the summary of the paper and our concluding remarks.    

\section{Methodology}
\label{sec:meth}
\subsection{Gutenberg-Richter power law}

The Gutenberg-Richter (GR) power law is one of the important empirical law that is applied to study the statistical properties of an earthquake in geophysics~\cite{sornette1999general,gutenberg1944frequency}. It shows the relationship between the number of earthquakes (aftershocks) to the magnitude of each earthquake. It states that the cumulative number of earthquakes ($N$) with magnitude $M$ larger than or equal to $M$ is proportional to $M$ and is given by

\begin{equation}
\label{eq:gr}
 \log_{10}N(M)=\alpha-\beta M 
\end{equation}

Where $\alpha$ and $\beta$ are two positive constants. The slope $\beta$ shows the relationship between the convergence process after a shock to its magnitude. 

Eqn.~\ref{eq:gr} is also used to explain the aftershock sequence in the stock market after a major market crash. The large absolute value of $\beta$ indicates that the stock/index will experience a heavy-sized aftershocks but will get to a normal period very shortly and low absolute value of $\beta$ implies that the stock/index will in general have a longer recovery period. The constant parameter $\alpha$ represents the number of remaining aftershocks, which does not depend on the magnitude of the aftershocks~\cite{siokis2012stock,kapopoulos2005stock}. In our case, magnitude (M) of the mainshock and aftershock is defined as the difference in the stock price from the day of fall till there is strong sign of recovery i.e., recovery of 7.0\% or more of the main crash.

\subsection{Degree of Nonstationarity}
\label{subsec:DNS}

In order to test nonstationarity, Degree of nonstationarity (DNS) test is applied. DNS is defined as~\cite{chowdhury2017identification,huang1998empirical,mahata2021characteristics}

\begin{equation}
DNS(\omega)=\frac{1}{T}\int_{0}^{T}[1-\frac{H(t,\omega)}{h(\omega)}]dt
\label{eqn:dns}
\end{equation}

Here [0, $T$] is the time window, $h(\omega)$ is the marginal spectrum which is written as $h(\omega)=\frac{1}{T}\int_{0}^{T}H(t,\omega) dt$ and $H(t,\omega)$ is the Hilbert spectrum. Hilbert spectrum is defined as $H(t,\omega)=Re\sum_{j} M_j(t)~e^{i\int{\omega_j(t)dt}}$, where $M_j(t)$ is the amplitude and $\omega_j(t)$ is the frequency. 

Eqn.~\ref{eqn:dns} shows the stationarity of a time series data as a function of frequency. In case of a stationary data, Hilbert spectrum is independent of time and will consist of horizontal line. Hence, $DNS(\omega)$ will be zero. However, when the $H(t,\omega)$ is time dependent $DNS(\omega)$ will not be zero~\cite{chowdhury2017identification,huang1998empirical,mahata2021characteristics}.

\subsection{Structural Break Analysis Technique}
\label{subsec:sba}
The structural break analysis method developed by Bai and Perron (1998, 2003)~\cite{bai1998estimating,bai2003computation} is applied to capture the structural changes in a stock price time series.  As the structural breaks emerge for a sudden change in a time series, it helps us to identify the duration of the influence of a crisis~\cite{ bai1998estimating,bai2003computation,MAHATA2020123612}.
A detailed procedure of the estimation of structural break is given below~\cite{bai1998estimating,bai2003computation}.

Let us consider the following structural change model with $m$ break points
\begin{equation}
 y_t=x_t^{'}\zeta+z_t^{'}\delta_j+u_t, ~~~   t=T_{j-1}+1,.....,T_j 
\end{equation}
where, $j=1,...,~m$. $x_t (p\times1)$ and $z_t(q\times1)$ are vectors of covariates and $u_t$ is the disturbance at time t.  The break dates ($T_1, . . . , T_m$) are explicitly treated as unknown. The purpose is to estimate the unknown regression coefficients $\zeta$, $\delta_j$ and the break dates. For each $m$-partition ($T_1, . . . , T_m$), denoted by {$T_j$},  the associate least square estimates of $\zeta$, $\delta_j$ are obtained by minimizing the  following objective function 
 \begin{equation}
 \label{eq:3}
 \sum_{i=1}^{m+1}\sum_{t=T_{i-1}+1}^{T_i}(y_t-x_t^{'}\zeta-z_t^{'}\delta_i)^2
 \end{equation}
 Let  $\hat\zeta(T_j)$ and $\hat\delta(T_j)$ denote the resulting estimates that minimize the objective function.  The sum of squared residual, $SSR_T(T_1,....,T_m)$ is obtained by substituting $\hat\zeta(T_j)$  and  $\hat\delta(T_j)$  in the above objective function. The resulting estimated break dates $T_1,....,T_m$ are obtained as
 \begin{equation}
 \{(\hat{T_1}),...,(\hat{T_m})\}=arg \min_{T_1,...T_m} SSR_T\{T_1,....,T_m\} 
\end{equation}
where the minimization is done over all partitions $(T_{1}, . . . , T_{m})$.  The break date estimators are global minimizer of the objective function. Let $v(i,j)$ be the recursive residual at time $j$ obtained using a sample that starts at date $i$. $SSR(i,j)$ is the sum of squared residual obtained by applying least-squares to a segment that starts at date $i$ and ends at date $j$ with $SSR(i,j)=SSR(i,j-1)+v(i,j)^{2}$. In order to evaluate the global optimal partition, a dynamical programming approach is applied on the sums of squared residual of all the relevant combination of ($i,j$) segments. Finally, $m$ break points have been obtained from the global optimal partition function $SSR(T_{m,K})$ by using first $K$ observation as given in the following recursive equation.
 
\begin{equation}
SSR(T_{m,K})= \min_{mh\leq K \leq K-h} \{SSR(T_{m-1,j})+SSR(j+1,K)\}
\end{equation}
In our regression procedure, we have taken upto 3-break points or four segments and $z_t=\{1\}$ as a constant regressor. 

\section{Data Analysed}
\label{sec:da}
In this paper, analysis of the stock price crashes are carried out using the daily closing price of indices and companies. The stock data were taken from Yahoo Finance~\cite{Yahoofi}. We have analyzed the shocks due to the 1987 crash, the financial-crisis-2008 and the 2020 COVID-19 pandemic. In case of the 1987 crash, we have taken DJIA index data as daily closing price of other companies were unavailable.

 We have taken fourteen dominant world stock indices: Nasdaq (USA), DJIA (USA), HSI (Hongkong), BEL20 (Belgium), IBOVESPA (Brazil), BSE SENSEX (India), DAX (Germany), IBEX35 (Spain), IPC (Mexico), Nifty50 (India), NIKKEI225 (Japan), S\&P500 (USA), SSE (China), CAC40 (France) and five Indian sectoral indices. The reason for choosing these indices are that they broadly represent the activity of world stock markets. Further, we have also taken thirty-two important companies that are listed in different leading worldwide stock exchanges.
%Further, we have also taken thirty-seven listed companies and sectoral indices under these world indices. 

\section{Results and Discussion}
\label{sec:res}
 
In this section, we have presented the results of analysis of  the mainshock and its subsequent aftershocks of the 1987 crash, the financial-crisis-2008 and the 2020 COVID-19 pandemic using our proposed definition of market crash. We have calculated the influence time length of these crashes using 3-point structural break analysis. All the stocks mentioned in Sec.~\ref{sec:da} have been analyzed and their results are discussed herewith.

\begin{figure}
\center
\includegraphics[width=8cm]{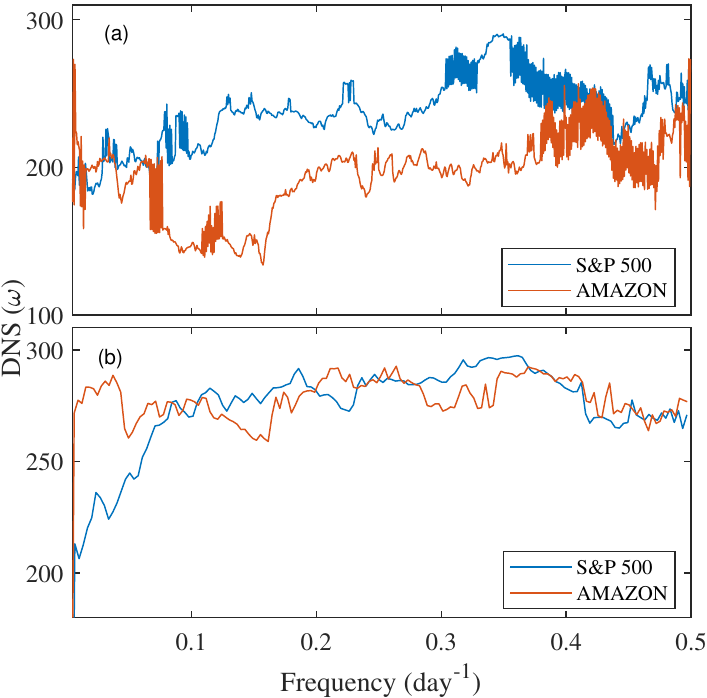}
\caption{\label{fig:DNS}It represents the DNS plot of S\&P 500 index and Amazon Inc. during (a) the financial-crisis-2008 and (b)  the 2020 COVID-19 pandemic. The plots clearly show that the stock prices are non-stationary.}
\end{figure}

We have carried out the degree of non-stationarity (DNS) test for all the stocks and indices. Fig.~\ref{fig:DNS} (a) and (b) represent the DNS($\omega$) plot of S\&P 500 index and Amazon Inc. during the financial-crisis-2008 and the 2020 COVID-19 pandemic, respectively.  DNS($\omega$) shows whether the time series data is stationary or not. DNS vs. $\omega$ plot is a horizontal straight line for a stationary time series. The DNS($\omega$) is not a horizontal straight line for nonstationary time series. Fig.~\ref{fig:DNS} (a) and (b) show that the time series data are nonstationary as DNS vs $\omega$ plot are not a straight line. The DNS test show that the remaining indices and companies are also nonstationary. As the time series data are nonstationary, the structural break analysis is carried out to estimate the crash and recovery points.

\subsection{The crash of 1987}
\begin{figure}
\center
\includegraphics[width=8cm]{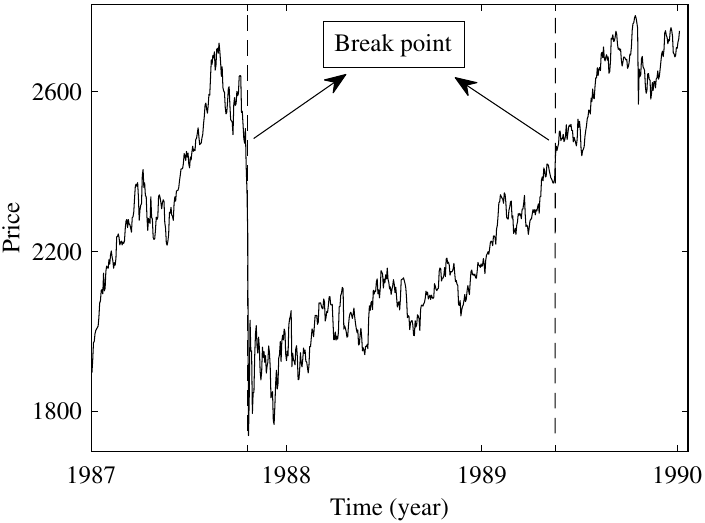}
\caption{\label{fig:1987D}It represents the daily closing price of DJIA index from January 1987 to December 1989. The vertical dotted lines represent the structural break points.}
\end{figure}

We first investigate the DJIA index from 1987 to 1990 during and after the Black Monday crash occurred at New York Stock Exchange (NYSE) that was considered one of the worst crash in the history of the NYSE. We found that the crash continued for eleven consecutive days. Hence, this consecutive eleven days fall is the actual crash-time and should be considered as a single crash. Fig.~\ref{fig:1987D} represents the daily closing price of DJIA index during that period. In order to identify the number of aftershocks due to the mainshock of the 1987 crash, the duration of the influence is identified using the structural break analysis as given in Subsec.~\ref{subsec:sba}. The closest breakpoints are shown by the vertical dotted lines in Fig.~\ref{fig:1987D}. The first vertical dotted line represents the starting point of the mainshock, and the second dotted line represents the end of the effect of the mainshock. Hence, the time of influence of the 1987 crash is from the time of the main crash till the end of 1989. The identified period is consistent with the recovery of the market.

\begin{figure}
\center
\includegraphics[width=8cm]{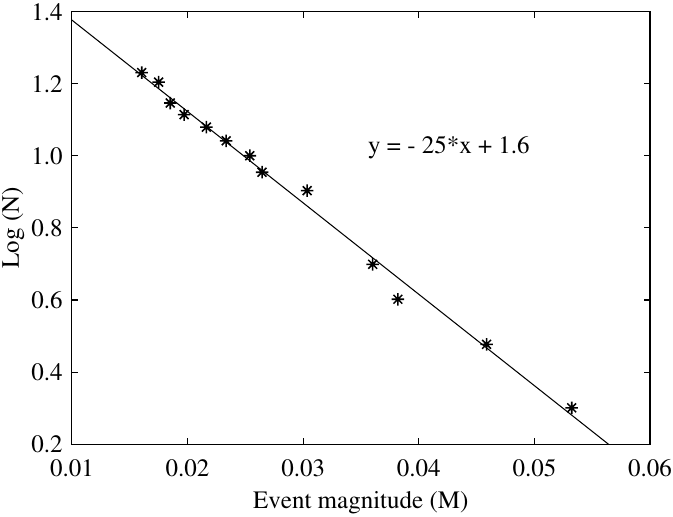}
\caption{\label{fig:1987GR}The figure represents the plot between cumulative number of events ($N$) and the corresponding magnitude ($M$) of the 1987 crash.}
\end{figure}

\begin{figure}
\center
\includegraphics[width=8cm]{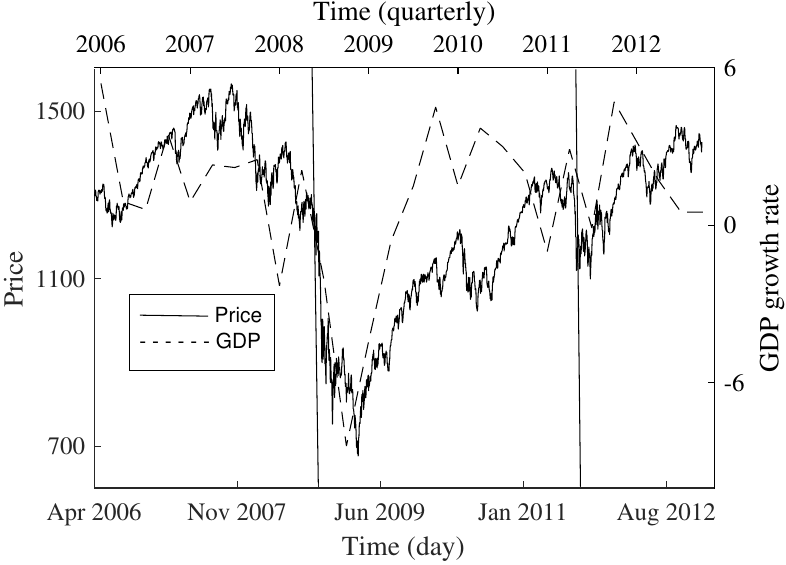}
\caption{\label{fig:GDPFC}Solid line represents the daily closing price of S\&P500 index from April 2006 to December 2012. The dashed line represents the quarterly GDP growth rate of USA from 2006 to 2012. The vertical lines represent the structural break points.}
\end{figure}

Fig.~\ref{fig:1987GR} shows the log-log plot between cumulative number of events (N) and its corresponding magnitude (M) of DJIA index during the 1987 crash. We performed a fit with the functional form of Eq.~\ref{eq:gr}. The typical plot of $\log(N)$ vs. $\log(M)$ describes the empirical data well for the study period, i.e., the 1987 crash and its aftershocks follow the GR power law. The straight line represents the best fit, and the values of $\alpha$ and $\beta$ mentioned in Eqn.~\ref{eq:gr} are 1.6 and 25.0, respectively. The values of $R^2$ and P-value are 0.988 and 0.998, respectively.  The figure clearly shows that the aftershocks with high magnitude are comparatively less than aftershocks with low magnitude. This should be the case as index generally does not fall with a high magnitude regularly. We have also obtained similar paradigmatic behavior of the leading stock price indices and companies due to the financial-crisis-2008 and the 2020 COVID-19 pandemic, and the detailed analysis are given below.

\subsection{The financial-crisis-2008}
In Fig.~\ref{fig:GDPFC}, the solid line and dashed line represent the daily closing price of S\&P 500 index from 2006 to 2012 and the quarterly GDP growth rate of USA from 2006 to 2012, respectively. The vertical solid lines are the structural breaks obtained for the financial-crisis-2008. The financial crisis can be clearly captured from these solid vertical lines. The closest breakpoints of the mainshock and recovery are mid of 2008 and end of 2011, respectively.  Hence, we have taken the data from the time of mainshock till the end of 2011 to analyze the influence of the mainshock. Similar analyses have been carried out for the rest of the 50 companies and indices. Detailed analysis of all the indices and companies are given in Table~\ref{tab:1}.   

\begin{table*}

\tiny
{\begin{tabular}{|l|c|c|c|c|c|c|c|c|c|c|c|c|}
\hline
Index/Company & \multicolumn{6}{c|}{The financial-crisis-2008} & \multicolumn{6}{c|}{The COVID-19 Crisis}  \\
\cline{2-13}
  &\% fall& N.A.S  & $\alpha$ & $\beta$ & P-value&$R^{2}$& \%fall &N.A.S   & $\alpha$& $\beta$&P-value&$R^{2}  $  \\
\hline
   NCI &22   & 26    &1.7   &14.0 & 0.892& 0.987 &14  &19      &1.6   &39.0 &0.741 &0.917   \\
\hline
    HSI&29  & 29    &1.7   &13.0 &0.741 & 0.923 &15 & 13    &1.4   &28.0  &0.995 &0.933  \\
\hline
    BEL20 Index&24  & 27    &1.7   &13.0 &0.995 &0.985 &30   & 17     &1.5   &30.0   &0.930 &0.946  \\
\hline
    IBOVESPA Index &29 & 26     &1.6   &9.9  &0.892 & 0.967  &21& 16    &1.4   &17.0   &0.912 &0.969  \\
\hline
    BSE Index&25  & 22     &1.7   &13.0  &0.979 &0.964 &20  &  13    &1.2   &12.0   &0.828 &0.900 \\
\hline
     DAX Index &27&27    &1.6   &11.0  &0.698 &0.967 &28 & 16  &1.5   &34.0  &0.912 &0.946   \\
\hline
   DJIA Index &22 & 21    &1.6   &14.0  &0.797 &0.935 &15  &18    &1.3   & 17.0  &0.709 &0.938 \\
\hline
 IBEX35 Index&25 & 22    &1.7   & 13.0  &0.979 & 0.964&20  & 13     &1.2   &12.0   &0.828 &0.894  \\
\hline
  IPC Index&26& 24    &1.3  &8.2  &0.953 &0.977  &13  &12    &1.3   & 26.0   &0.991 &0.970 \\
\hline
 Nifty50 Index&22 &18   &1.7   &14.0   &0.998 &0.962  &17& 11     &1.3   &24.0   &0.985 &0.940  \\
\hline
  NIKKEI225 Index&32&26     &1.7   &12.0  &0.992 &0.987  &29&13     &1.2   & 26.0  &0.433 &0.967  \\
\hline
 S\&P500 Index&23 & 27    &1.7   &13.0  &0.698 &0.966 &14  & 13   &  1.3 & 17.0  &0.995 &0.944 \\
\hline
 SSE Index &26& 18    & 1.5  &9.7  & 0.945&0.987 &12  & 11   &1.2   & 19.0   &0.991 &0.909 \\
\hline
CAC40 Index &27 &34    &1.8   &13.0   &0.962 &0.984 &29 &14      &1.5   &28.0   &0.862 & 0.918 \\
\hline
Nifty Realty  &41 &51   &1.9   &9.2 &0.850  &0.975   & 19   &15&1.3   &12.0 &0.998&0.917 \\
\hline
Nifty Pharma&26   &11 &1.3   &9.7   &0.985 &0.931  &15   &11    &1.3&22.0   &0.985 &0.919 \\
\hline
Nifty IT&22   &28 & 1.8  &15.0   & 0.490&0.989  &5   &12    &1.3&35.0   &0.433 &0.966 \\
\hline
Nifty Bank&32   &44 &1.8 &9.8  & 0.777  &0.975 &33  &17   &1.5    &19.0&0.998 &0.917 \\
\hline
Nifty FMCG&21   &19 &1.5   &15.0   &0.462 &0.973  &23   &8    &1.1& 15.0  &0.929 &0.903 \\
\hline
 Adidas AG&29& 32     &1.8   & 14.0  & 0.951&0.971 &32& 18  &1.4   &19.0   &0.945 &0.950  \\
\hline
  Amazon Inc.&29& 39    &1.9   & 15.0  &0.981 &0.985 &13&23  &1.5   & 22.0   &0.842 &0.980 \\
\hline
  3M Co.&26& 31    & 1.7  & 14.0  & 0.56 &0.945 & 14&15    & 1.4  & 23.0   &0.998 &0.925 \\
\hline
  
  BASF SE&31& 36     & 1.8  & 12.0   & 0.971& 0.969&28&19     &1.5   &19.0   &0.956 &0.989  \\
\hline

 TCS Ltd.&23 & 39     & 2.0  & 18.0  & 0.884& 0.969 &17 &  19   & 1.5  & 26.0   &0.956 &0.962 \\
\hline
  
  Daikin Industries&43&48     & 1.9  &10.0    & 0.480& 0950&24 &11   &1.4   &26.0    &0.985 &0.888 \\
\hline
  Bajaj Fin. Ltd.&48& 40    &1.9   &13.0  &0.983 &0.980 &23 & 22     &1.5   & 13.0   &0.990 &0.970 \\
\hline
  Braskem&43 &38     &1.7   &8.7    &0.693 &0.959 &49 & 21     &1.5   & 13.0   & 0.531& 0.960\\
\hline
 BPCL&26 & 38    &1.8   & 11.0 & 0.978 &0.990 & 20 &16     &1.3   & 12.0  &0.912 &0.969  \\
\hline
 Canon Inc.&26 & 31    & 1.9  &17.0   &0.998  &0.985 &  25& 18    &1.5   & 22.0   &0.673 &0.940 \\
\hline
  CCCL&33&45     &1.9   &11.0   &0.929 &0.992 &24  &17    &1.5   &19.0   &0.673 &0.970  \\
\hline
  HDFC Bank Ltd.&26&  27   &1.8   &16.0    &0.905 &0.977  &29 & 17   & 1.4  &17.0   &0.930 &0.979  \\
\hline
  Infosys Ltd.&25& 29    &1.8   & 14.0  & 0.72& 0.98&30 & 18  &1.3   & 22.0   &0.995 &0.993 \\
\hline
 Intel Corp.&27 &39   &2.0   &17.0 &0.884 &0.975 &10  &20    &1.5   &24.0   &0.956 &0.954  \\
\hline
 Microsoft Corp.&26 & 27    &1.7   &14.0 & 0.905& 0.940&16  &23     &1.5   &16.0  &0.593 &0.929  \\
\hline
THL&29 &  50    &2  & 17.0 &0.639 &0.979 &16 &19  & 1.6  & 31.0   &0.998 &0.970 \\
\hline
  BATP&25& 23   &1.7   &18.0 &0.983 &0.984   &20&  14   &1.3   & 17.0   &0.828 &0.827 \\
\hline
 
 Advantest Corp.&47 & 47   &1.9   &9.5 &0.935 &0.980 &18  & 17    &1.4   &13.0    &0.633 &0.920 \\
\hline

  BMW AG &37&48 & 2.0  &14.0  & 0.995& 0.985 &17  & 17     &1.6   &24.0   &0.930 &0.960  \\
\hline
  HII&37 &27   &1.6   & 8.6  &0.698 &0.893 &30& 19    &1.5   &22.0    &0.956 &0.970 \\
\hline
 Home Depot Inc.&25 & 36  &  1.8 & 13.0 &0.851 & 0.973 &20& 14    & 1.4  & 21.0   &0.862 &0.986 \\
\hline
Daiichi Sankyo Com&45 & 24   & 1.6  & 9.3 & 0.861& 0.975&19 &11  &1.1  &13.0     &0.736 &0.911\\
\hline  
Apple Inc.&  28 &39      &1.8   &13.0 &0.514 & 0.986      &17   &18  &1.4  &14.0   &0.945 &0.950 \\
\hline
 
 GlaxoSmithkline plc&23   &22   &1.6   &14.0    &0.821 &0.930   &16   &11    &1.3&21.0   &0.985 &0.972 \\
\hline
 UCL& 34  &28   &1.7   &12.0   &0.905 &0.962    & 16  &13    &1.3&15.0   & 0.995&0.889 \\
\hline
 Coca-Cola Co.&23   &20   &1.7   &20.0    &0.966 & 0.982  &20   &15    &1.5&29.0  &0.998 &0.958 \\
\hline
 PepsiCo, Inc.&21   &19  &1.6   &16.0   &0.998 &0.984    & 19  &10    &1.2&25.0   &0.675 &0.934 \\
\hline

SPGI &34   &53 &2   &13.0   &0.866 &0.954  &50   &25    &1.4& 8.0  &0.414 &0.980 \\
\hline
Prologis Inc.&59   &61 & 1.9  &7.6   &0.700 &0.974  &18   &20    &1.4&15.0   &0.771 &0.961 \\
\hline
CBRE Group Inc.&47   &73 &2   &8.5   &0.759 &0.987  &24   &18 &   1.4&12.0& 0.945  &0.966  \\
\hline
ACI &30   &46 &2   &15.0   &0.625 &0.983  &33   &21    &1.4&15.0   &0.797 &0.906 \\
\hline
ATC.& 38  &29 &1.8   &16.0   &0.927 &0.977  &18   &18    &1.5& 23.0  &0.709 &0.953 \\
\hline

\end{tabular}
\label{tab:1}}
\caption{\tiny It shows the analysis of 51 indices and companies. Col-2 and col-3 under the financial-crisis-2008 and the COVID-19 Crisis represent \% of fall of the mainshock and the number of aftershocks(N.A.S) respectively. Other parameters have usual meanings. Abrebiation of a few companies:- Nasdaq Composite Index: NCI; Hang Seng Index: HSI; Bharat Petroleum Corp. Ltd:BPCL; Casio Computer Co. Ltd:CCCL; Tencent Holdings Ltd: THL; British American Tobacoo plc: BATP; Honeywell International Inc.: HII; Simon Property Group Inc.:SPGI; AvalonBay Communitites Inc:ACI; Ultratech cement Ltd:UCL; American Tower Corporation: ATC.}
\end{table*}

\begin{figure}
\center
\includegraphics[width=8cm]{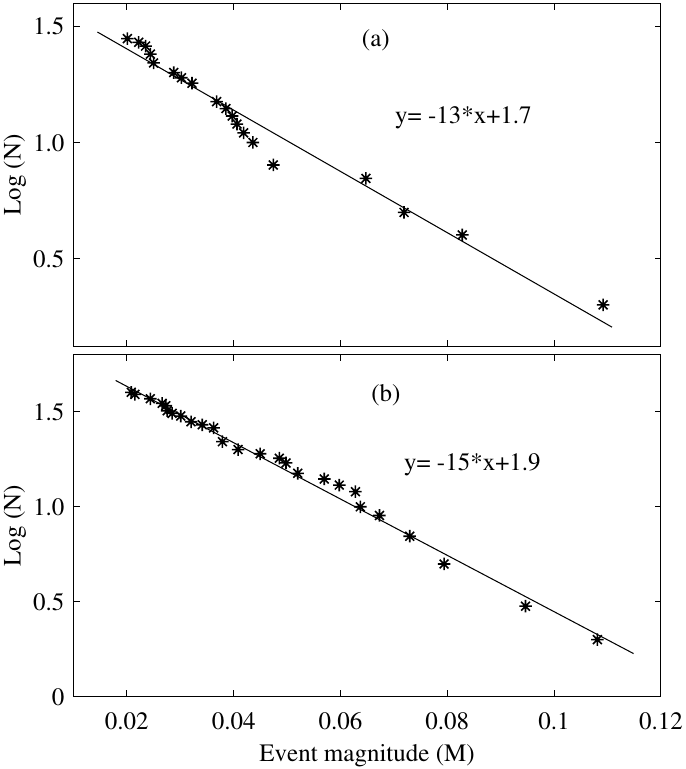}
\caption{\label{fig:GRFC}It represents the plot between cumulative number of events ($N$) and the corresponding magnitude ($M$) of (a) S\&P500 index and (b) Amazon Inc., respectively during the financial-crisis-2008.}
\end{figure}

Fig.~\ref{fig:GRFC} (a) and (b) show the typical plot of $\log(N)$ versus $\log(M)$ of S\&P 500 index and Amazon Inc., respectively during the financial-crisis-2008. The points show the aftershocks with its magnitude, and the straight line represents the best fit of the aftershock sequence. We clearly see that the number of aftershocks with high magnitude is very low, and the number of aftershocks with low magnitude is very high which can be observed from the cluster of points formed around low magnitude region. This shows that the aftershocks with high magnitude are rare and aftershocks with low magnitude occurs frequently until the market fully recovers from the shock. The obtained value of  $\alpha$ and $\beta$ of Eqn.\ref{eq:gr} during the financial crisis for S\&P 500 are 1.7 and 13.0 and for Amazon Inc. are 1.9 and 15.0, respectively. The absolute value of $\beta$ in Fig.~\ref{fig:GRFC} (a) is less than Fig.~\ref{fig:GRFC} (b) which indicates the later will experience large aftershocks with high volatility whereas the former will take longer time to recover. This is generally true as indices fall and recover slowly compared to individual stocks. The values of  $\alpha$ and $\beta$ are calculated for all the 51 stocks which is shown in Table~\ref{tab:1}. Adjusted $R^2$ and P-value are also calculated for the same. From the figure and the values of $R^2$ and P-value it is clear that aftershocks sequence of S\&P 500 and Amazon Inc. follow the GR power law during the financial-crisis-2008. Similar results have been found for the rest of the 49 companies for financial-crisis-2008. Detailed analysis of all the indices and companies are given in Table~\ref{tab:1}. 

\begin{figure}
\center
\includegraphics[width=8cm]{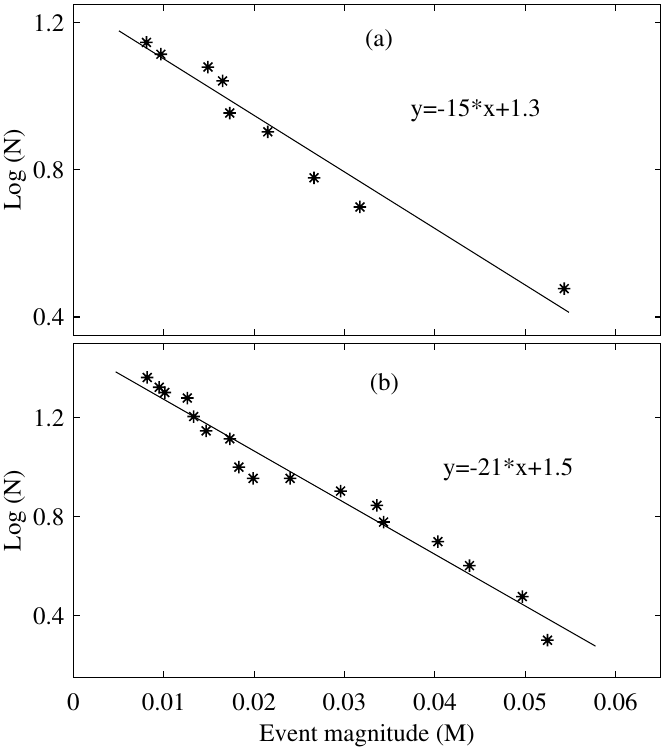}
\caption{\label{fig:GRCOVID}It represents the plot between cumulative number of events ($N$) and the corresponding magnitude ($M$) of (a) S\&P500 index and (b) Amazon Inc., respectively during the COVID-19.}
\end{figure}

\subsection{The COVID-19 pandemic}
Fig.~\ref{fig:GRCOVID} (a) and (b) show the plots of $\log(N)$ vs. $\log(M)$ of S\&P 500 index and Amazon Inc., respectively during the ongoing COVID-19 pandemic. The straight line represents the best fit and the values of $\alpha$ and $\beta$ for S\&P 500 are 1.3 and 15.0 and for Amazon Inc. are 1.5 and 21.0, respectively. It is observed that the absolute value of $\beta$ in Fig.~\ref{fig:GRCOVID} (a) is less than Fig.~\ref{fig:GRCOVID} (b). This imply that the Amazon Inc. has higher volatility and S\&P 500 index will have a longer turbulent period. It is also clear from Fig.~\ref{fig:GRCOVID}(a) and (b) that the aftershock sequence of S\&P 500 and Amazon Inc. follow the GR power law during this ongoing pandemic. We have carried out the analysis for all the 51 companies and indices and found that these indices and companies follow the GR power law. Detailed analysis of all the indices and companies are given in Table~\ref{tab:1}. 

It is interesting to note that for the COVID-19 pandemic, there is no cluster of points in the low magnitude region as observed in the Fig.~\ref{fig:GRFC} (a) and (b), and a gap between the highest magnitude aftershock and its next aftershock is observed in Fig.~\ref{fig:GRCOVID} (a). These observations give us a strong indication that more aftershocks with low magnitude can be anticipated in the near future due to the ongoing COVID-19 pandemic. The gap seen in high magnitude region in Fig.~\ref{fig:GRCOVID} (a) shows the possibility of high-magnitude aftershocks in the near future.

The results show that the absolute $\beta$ is greater for the COVID-19 pandemic than the financial-crisis-2008. This implies that the stock market was highly volatile during the COVID-19 pandemic but its recovery was also swift. Whereas, the financial-crisis-2008 had a much severe impact on the stock market in terms of duration of its impact which lasted for few years. The difference in $\beta$ value indicates that the effect of the COVID-19 pandemic will not last longer than the financial-crisis-2008.

\begin{figure}
\center
\includegraphics[width=8cm]{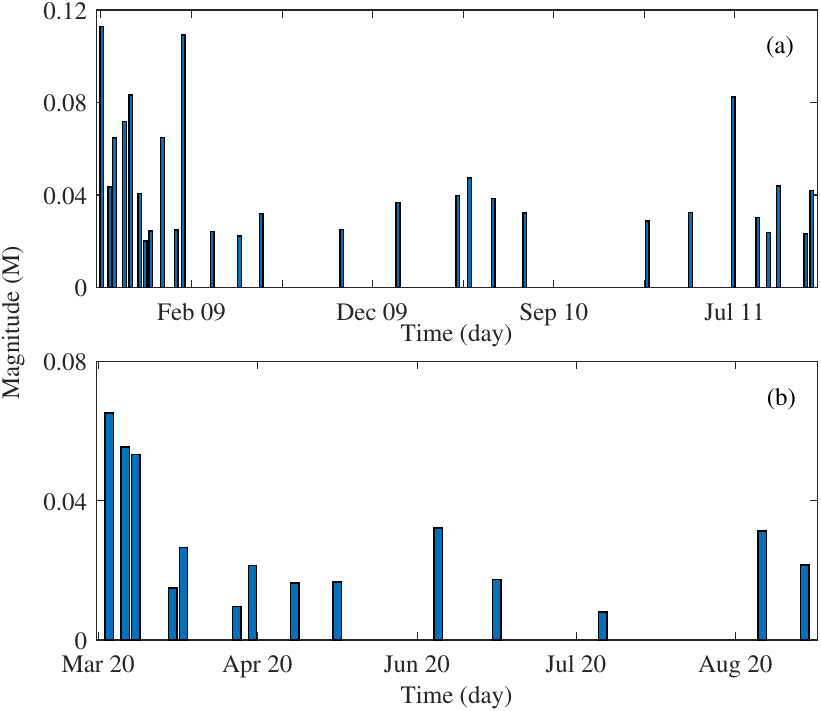}
\caption{\label{fig:spaf}It represent the mainshock and the aftershocks of S\&P 500 index during (a) the financial-crisis-2008 and (b) the COVID-19 pandemic, respectively.}
\end{figure}

\subsection{Analysis of temporal variance and Inter-occurrence time of aftershocks}
The temporal variance of the aftershock sequence is presented in Fig.~\ref{fig:spaf} (a) and (b) of S\&P 500 during the financial-crisis-2008 and the COVID-19 pandemic, respectively. They show that the number of aftershocks per unit time does not decay with time. Sometimes the number of aftershocks per unit time increases and hence does not follow the Omori law. This finding is in contradiction to the previous studies~\cite{siokis2012stock,selccuk2004financial,petersen2010market}. Our finding is consistent with the statistics of the mainshock and aftershocks in stock price as explained below.  

After a major crash, the market recovers for a certain period, but due to some negative perturbation, stressed condition of economy and fear of investors lead to smaller crashes. During the stressed period of the market recovery, it is observed that the number of aftershocks per unit time increases significantly at the later period of market recovery. Hence, it is unlikely to follow Omori law during the market recovery after a major crash. The studies shown in Ref.~\cite{siokis2012stock,selccuk2004financial,petersen2010market} on the decay of aftershocks with time that holds Omori law may be incomplete due to the erroneous definition of the mainshock, aftershocks and duration of influence that is discussed in Sec.\ref{sec:intro}.   

\begin{figure}
\center
\includegraphics[width=8cm]{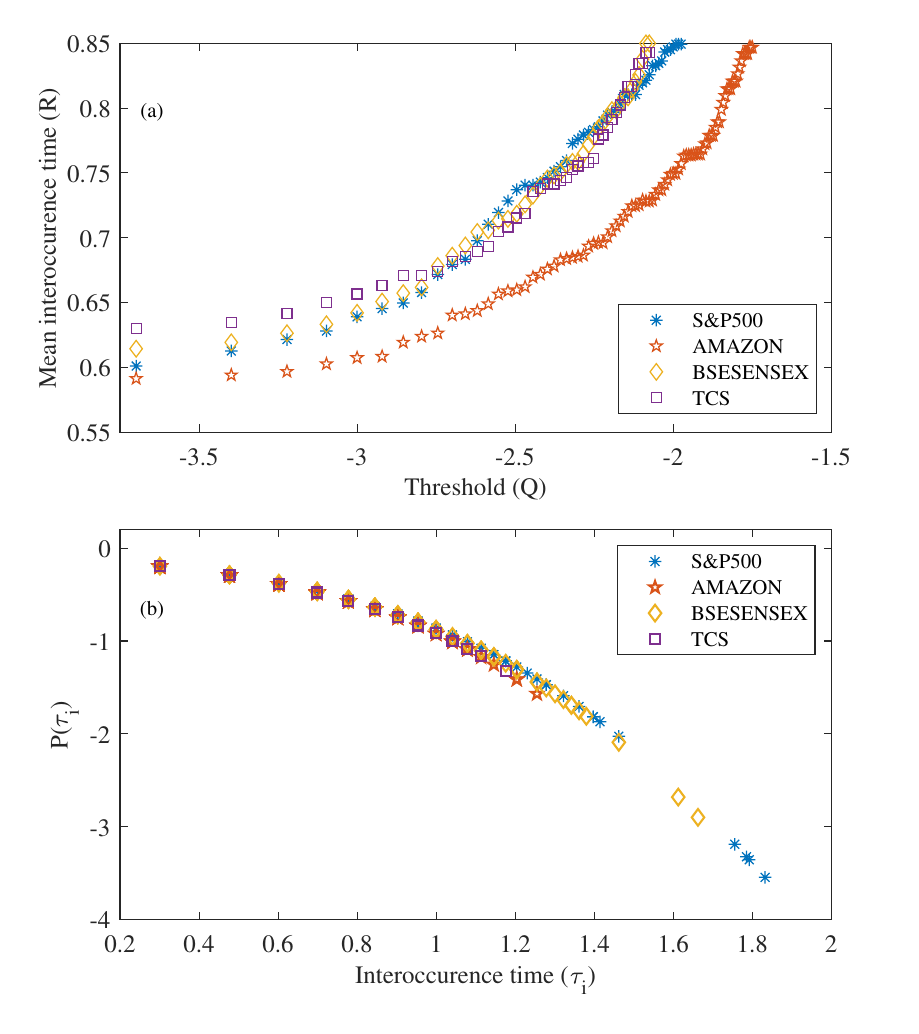}
\caption{\label{fig:intero}It represents the plot between (a) mean occurrence time ($R$) and threshold ($Q$) and, (b) the distribution $P_Q (r)$ and interoccurence time ($r$), for S\&P 500, Amazon Inc., BSE SENSEX and TCS Ltd.}
\end{figure} 

\begin{figure}
\center
\includegraphics[width=8cm]{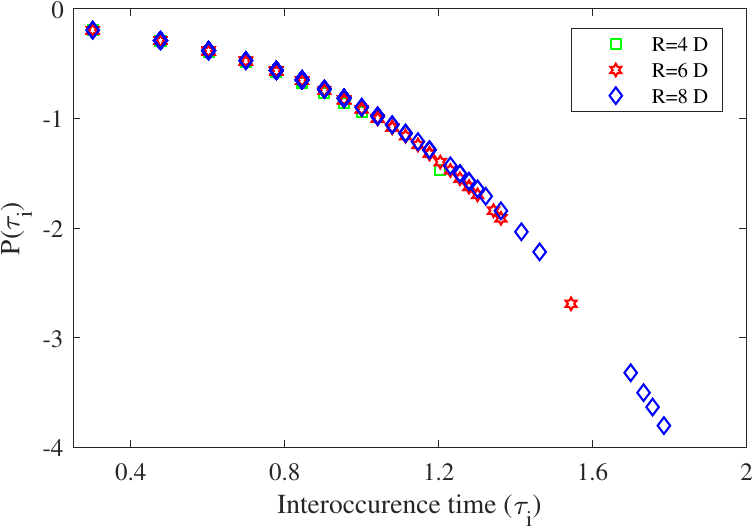}
\caption{\label{fig:intero2}It represents the plot between the distribution $P(\tau_i)$ and occurrence time $(\tau_i)$ of S\&P 500 index for different mean interoccurence time ($R$)}
\end{figure} 

To study the characteristics of the inter-occurrence time($\tau_i$) between the aftershocks above a threshold ($Q$), we analyze the distribution P($\tau_i$). We found that for all the stocks, the $P(\tau_i)$ follows the generalized Pareto distribution~\cite{reiss1997statistical}, i.e., $\displaystyle~P(\tau_i)\propto\frac{1}{\{1+\lambda(q-1)\tau_i\}^{\frac{1}{(q-1)}}}$, where $\lambda$ and $q$ are constants. $q$ can be estimated using relation $\displaystyle~q=1+q_0+\ln{\frac{R}{2}},$ where $R$ is the mean inter-occurence time. Fig.~\ref{fig:intero} (a) shows the plot of $R$ vs. $Q$ for S\&P 500 index, Amazon Inc., BSE SENSEX index and TCS Ltd. during the financial-crisis-2008. The plot shows that the value of $R$ increases slowly at low value of $Q$ but for higher $Q$ values $R$ increases by power law. Fig.~\ref{fig:intero} (b) shows the plot of $P(\tau_i)$ vs. $\tau_i$ for S\&P 500 index, Amazon Inc., BSE SENSEX index and TCS Ltd. Earlier, we had considered any fall that is greater than 7\% of the main shock as an aftershock. Hence, in this analysis we have given a maximum $Q$ of 7\% of the mainshock for all the stocks and indices. We have obtained the value of $R$ for all the stocks. As all the $R's$ are in very close range, the variation in $P(\tau_i)$ with respect to $R$ is negligible. Fig~\ref{fig:intero2} shows the plot of $P(\tau_i)$ vs. $(\tau_i)$ of S\&P 500 index for different $R$ i.e., 4, 6, and 8 trading days which is seen to overlap with each other.

\section{Conclusion}
\label{sec:con}
This paper draws the limitation of the previous works with respect to their interpretation of crash and duration of effect of the mainshock. The statistical nature of the aftershock sequence after major stock market crashes for the 1987 crash, the financial-crisis-2008 and the ongoing COVID-19 pandemic are analysed using the new definition of crash. The analysis is carried out for 51 stocks and indices, considering the mainshock or aftershock as a cumulative consecutive fall of the stock price. The duration of influence of the mainshock is obtained by the 3-point structural break method. The mainshock and aftershocks follow the Gutenberg-Richter power law. We found that in the 1987 crash and the financial-crisis-2008 there are a dense cluster of aftershocks in the low magnitude region. This indicates that aftershock of low magnitude occurs more frequently rather than high magnitude aftershock. 

The analysis shows that in the ongoing COVID-19 event, the number of aftershocks in the low-magnitude region is less, and there is a gap in the high-magnitude region. These results indicate that few more aftershocks with smaller amplitudes may be anticipated in the near future during this ongoing pandemic till the pandemic is completely eliminated. However, the probability of an aftershock with a large magnitude during the ongoing COVID-19 pandemic is very less as it is no longer a novel risk to the market. On comparison of the $\beta$ values between the COVID-19 and the financial-crisis-2008, we conclude that the effect of the COVID-19 will not last as long as the financial-crisis-2008. The analysis shows that the number of aftershocks does not necessarily decay with time, and hence, does not obey Omori's power law. However, inter-occurence-time of the aftershocks follows the generalized pareto distribution. 

Our finding is also consistent with the psychology of the investors that when the unique crisis becomes known, the market does not react too irrationally as it did initially, and hence subsequent crashes becomes relatively smaller. When smaller crashes happen during the recovery period, investors need not be worried of such smaller crashes as these crashes will be recovered very quickly. Further, these smaller crashes give the investors opportunities to enter in the stock market. The analysis may help investors make rational investment decisions during the stressed period after a major market crash. 
\section*{Acknowledgment}
NIT Sikkim is appreciated for allocating doctoral research fellowships to A.R. and A.M.
We appreciate the help of Sandeep Parajuli in this work. We also appreciate the comments from anonymous reviewer to improve the manuscript.

%\begin{thebibliography}{000} %for 3 digits
%\begin{thebibliography}{00}  %for 2 digits
%\begin{thebibliography}{0}   %for 1 digit

%\bibliographystyle{ws-ijmpc}
\bibliography{ref}

%merlin.mbs aipnum4-1.bst 2010-07-25 4.21a (PWD, AO, DPC) hacked
%Control: key (0)
%Control: author (8) initials jnrlst
%Control: editor formatted (1) identically to author
%Control: production of article title (0) allowed
%Control: page (1) range
%Control: year (1) truncated
%Control: production of eprint (0) enabled
\begin{thebibliography}{60}%
\makeatletter
\providecommand \@ifxundefined [1]{%
 \@ifx{#1\undefined}
}%
\providecommand \@ifnum [1]{%
 \ifnum #1\expandafter \@firstoftwo
 \else \expandafter \@secondoftwo
 \fi
}%
\providecommand \@ifx [1]{%
 \ifx #1\expandafter \@firstoftwo
 \else \expandafter \@secondoftwo
 \fi
}%
\providecommand \natexlab [1]{#1}%
\providecommand \enquote  [1]{``#1''}%
\providecommand \bibnamefont  [1]{#1}%
\providecommand \bibfnamefont [1]{#1}%
\providecommand \citenamefont [1]{#1}%
\providecommand \href@noop [0]{\@secondoftwo}%
\providecommand \href [0]{\begingroup \@sanitize@url \@href}%
\providecommand \@href[1]{\@@startlink{#1}\@@href}%
\providecommand \@@href[1]{\endgroup#1\@@endlink}%
\providecommand \@sanitize@url [0]{\catcode `\\12\catcode `\$12\catcode
  `\&12\catcode `\#12\catcode `\^12\catcode `\_12\catcode `\%12\relax}%
\providecommand \@@startlink[1]{}%
\providecommand \@@endlink[0]{}%
\providecommand \url  [0]{\begingroup\@sanitize@url \@url }%
\providecommand \@url [1]{\endgroup\@href {#1}{\urlprefix }}%
\providecommand \urlprefix  [0]{URL }%
\providecommand \Eprint [0]{\href }%
\providecommand \doibase [0]{http://dx.doi.org/}%
\providecommand \selectlanguage [0]{\@gobble}%
\providecommand \bibinfo  [0]{\@secondoftwo}%
\providecommand \bibfield  [0]{\@secondoftwo}%
\providecommand \translation [1]{[#1]}%
\providecommand \BibitemOpen [0]{}%
\providecommand \bibitemStop [0]{}%
\providecommand \bibitemNoStop [0]{.\EOS\space}%
\providecommand \EOS [0]{\spacefactor3000\relax}%
\providecommand \BibitemShut  [1]{\csname bibitem#1\endcsname}%
\let\auto@bib@innerbib\@empty
%</preamble>
\bibitem [{\citenamefont {Sornette}(2017)}]{sornette2017stock}%
  \BibitemOpen
  \bibfield  {author} {\bibinfo {author} {\bibfnamefont {D.}~\bibnamefont
  {Sornette}},\ }\href@noop {} {\emph {\bibinfo {title} {Why stock markets
  crash: critical events in complex financial systems}}},\ Vol.~\bibinfo
  {volume} {49}\ (\bibinfo  {publisher} {Princeton University Press},\ \bibinfo
  {year} {2017})\BibitemShut {NoStop}%
\bibitem [{\citenamefont {McNamara}\ \emph {et~al.}(2015)\citenamefont
  {McNamara}, \citenamefont {Benz}, \citenamefont {Herrmann}, \citenamefont
  {Bergman}, \citenamefont {Earle}, \citenamefont {Holland}, \citenamefont
  {Baldwin},\ and\ \citenamefont {Gassner}}]{mcnamara2015earthquake}%
  \BibitemOpen
  \bibfield  {author} {\bibinfo {author} {\bibfnamefont {D.~E.}\ \bibnamefont
  {McNamara}}, \bibinfo {author} {\bibfnamefont {H.~M.}\ \bibnamefont {Benz}},
  \bibinfo {author} {\bibfnamefont {R.~B.}\ \bibnamefont {Herrmann}}, \bibinfo
  {author} {\bibfnamefont {E.~A.}\ \bibnamefont {Bergman}}, \bibinfo {author}
  {\bibfnamefont {P.}~\bibnamefont {Earle}}, \bibinfo {author} {\bibfnamefont
  {A.}~\bibnamefont {Holland}}, \bibinfo {author} {\bibfnamefont
  {R.}~\bibnamefont {Baldwin}}, \ and\ \bibinfo {author} {\bibfnamefont
  {A.}~\bibnamefont {Gassner}},\ }\bibfield  {title} {\enquote {\bibinfo
  {title} {Earthquake hypocenters and focal mechanisms in central oklahoma
  reveal a complex system of reactivated subsurface strike-slip faulting},}\
  }\href@noop {} {\bibfield  {journal} {\bibinfo  {journal} {Geophysical
  Research Letters}\ }\textbf {\bibinfo {volume} {42}},\ \bibinfo {pages}
  {2742--2749} (\bibinfo {year} {2015})}\BibitemShut {NoStop}%
\bibitem [{\citenamefont {Sarlis}\ \emph {et~al.}(2018)\citenamefont {Sarlis},
  \citenamefont {Skordas}, \citenamefont {Varotsos}, \citenamefont
  {Ram{\'\i}rez-Rojas},\ and\ \citenamefont
  {Flores-M{\'a}rquez}}]{sarlis2018natural}%
  \BibitemOpen
  \bibfield  {author} {\bibinfo {author} {\bibfnamefont {N.~V.}\ \bibnamefont
  {Sarlis}}, \bibinfo {author} {\bibfnamefont {E.~S.}\ \bibnamefont {Skordas}},
  \bibinfo {author} {\bibfnamefont {P.~A.}\ \bibnamefont {Varotsos}}, \bibinfo
  {author} {\bibfnamefont {A.}~\bibnamefont {Ram{\'\i}rez-Rojas}}, \ and\
  \bibinfo {author} {\bibfnamefont {E.~L.}\ \bibnamefont
  {Flores-M{\'a}rquez}},\ }\bibfield  {title} {\enquote {\bibinfo {title}
  {Natural time analysis: On the deadly mexico m8. 2 earthquake on 7 september
  2017},}\ }\href@noop {} {\bibfield  {journal} {\bibinfo  {journal} {Physica
  A: Statistical Mechanics and its Applications}\ }\textbf {\bibinfo {volume}
  {506}},\ \bibinfo {pages} {625--634} (\bibinfo {year} {2018})}\BibitemShut
  {NoStop}%
\bibitem [{\citenamefont {Pisarenko}\ and\ \citenamefont
  {Sornette}(2003)}]{pisarenko2003characterization}%
  \BibitemOpen
  \bibfield  {author} {\bibinfo {author} {\bibfnamefont {V.}~\bibnamefont
  {Pisarenko}}\ and\ \bibinfo {author} {\bibfnamefont {D.}~\bibnamefont
  {Sornette}},\ }\bibfield  {title} {\enquote {\bibinfo {title}
  {Characterization of the frequency of extreme earthquake events by the
  generalized pareto distribution},}\ }\href@noop {} {\bibfield  {journal}
  {\bibinfo  {journal} {pure and applied geophysics}\ }\textbf {\bibinfo
  {volume} {160}},\ \bibinfo {pages} {2343--2364} (\bibinfo {year}
  {2003})}\BibitemShut {NoStop}%
\bibitem [{\citenamefont {Nguyen}, \citenamefont {Bhatti},\ and\ \citenamefont
  {Henry}(2017)}]{nguyen2017vietnam}%
  \BibitemOpen
  \bibfield  {author} {\bibinfo {author} {\bibfnamefont {C.}~\bibnamefont
  {Nguyen}}, \bibinfo {author} {\bibfnamefont {M.~I.}\ \bibnamefont {Bhatti}},
  \ and\ \bibinfo {author} {\bibfnamefont {D.}~\bibnamefont {Henry}},\
  }\bibfield  {title} {\enquote {\bibinfo {title} {Are vietnam and chinese
  stock markets out of the us contagion effect in extreme events?}}\
  }\href@noop {} {\bibfield  {journal} {\bibinfo  {journal} {Physica A:
  Statistical Mechanics and its Applications}\ }\textbf {\bibinfo {volume}
  {480}},\ \bibinfo {pages} {10--21} (\bibinfo {year} {2017})}\BibitemShut
  {NoStop}%
\bibitem [{\citenamefont {Piccoli}\ \emph {et~al.}(2017)\citenamefont
  {Piccoli}, \citenamefont {Chaudhury}, \citenamefont {Souza},\ and\
  \citenamefont {da~Silva}}]{piccoli2017stock}%
  \BibitemOpen
  \bibfield  {author} {\bibinfo {author} {\bibfnamefont {P.}~\bibnamefont
  {Piccoli}}, \bibinfo {author} {\bibfnamefont {M.}~\bibnamefont {Chaudhury}},
  \bibinfo {author} {\bibfnamefont {A.}~\bibnamefont {Souza}}, \ and\ \bibinfo
  {author} {\bibfnamefont {W.~V.}\ \bibnamefont {da~Silva}},\ }\bibfield
  {title} {\enquote {\bibinfo {title} {Stock overreaction to extreme market
  events},}\ }\href@noop {} {\bibfield  {journal} {\bibinfo  {journal} {The
  North American Journal of Economics and Finance}\ }\textbf {\bibinfo {volume}
  {41}},\ \bibinfo {pages} {97--111} (\bibinfo {year} {2017})}\BibitemShut
  {NoStop}%
\bibitem [{\citenamefont {Zhang}\ \emph {et~al.}(2009)\citenamefont {Zhang},
  \citenamefont {Yu}, \citenamefont {Wang},\ and\ \citenamefont
  {Lai}}]{zhang2009estimating}%
  \BibitemOpen
  \bibfield  {author} {\bibinfo {author} {\bibfnamefont {X.}~\bibnamefont
  {Zhang}}, \bibinfo {author} {\bibfnamefont {L.}~\bibnamefont {Yu}}, \bibinfo
  {author} {\bibfnamefont {S.}~\bibnamefont {Wang}}, \ and\ \bibinfo {author}
  {\bibfnamefont {K.~K.}\ \bibnamefont {Lai}},\ }\bibfield  {title} {\enquote
  {\bibinfo {title} {Estimating the impact of extreme events on crude oil
  price: An emd-based event analysis method},}\ }\href@noop {} {\bibfield
  {journal} {\bibinfo  {journal} {Energy Economics}\ }\textbf {\bibinfo
  {volume} {31}},\ \bibinfo {pages} {768--778} (\bibinfo {year}
  {2009})}\BibitemShut {NoStop}%
\bibitem [{\citenamefont {Sornette}\ and\ \citenamefont
  {Johansen}(1998)}]{sornette1998hierarchical}%
  \BibitemOpen
  \bibfield  {author} {\bibinfo {author} {\bibfnamefont {D.}~\bibnamefont
  {Sornette}}\ and\ \bibinfo {author} {\bibfnamefont {A.}~\bibnamefont
  {Johansen}},\ }\bibfield  {title} {\enquote {\bibinfo {title} {A hierarchical
  model of financial crashes},}\ }\href@noop {} {\bibfield  {journal} {\bibinfo
   {journal} {Physica A: Statistical Mechanics and its Applications}\ }\textbf
  {\bibinfo {volume} {261}},\ \bibinfo {pages} {581--598} (\bibinfo {year}
  {1998})}\BibitemShut {NoStop}%
\bibitem [{\citenamefont {Sornette}, \citenamefont {Johansen},\ and\
  \citenamefont {Bouchaud}(1996)}]{sornette1996stock}%
  \BibitemOpen
  \bibfield  {author} {\bibinfo {author} {\bibfnamefont {D.}~\bibnamefont
  {Sornette}}, \bibinfo {author} {\bibfnamefont {A.}~\bibnamefont {Johansen}},
  \ and\ \bibinfo {author} {\bibfnamefont {J.-P.}\ \bibnamefont {Bouchaud}},\
  }\bibfield  {title} {\enquote {\bibinfo {title} {Stock market crashes,
  precursors and replicas},}\ }\href@noop {} {\bibfield  {journal} {\bibinfo
  {journal} {Journal de Physique I}\ }\textbf {\bibinfo {volume} {6}},\
  \bibinfo {pages} {167--175} (\bibinfo {year} {1996})}\BibitemShut {NoStop}%
\bibitem [{\citenamefont {Mahata}\ \emph
  {et~al.}(2021{\natexlab{a}})\citenamefont {Mahata}, \citenamefont {Rai},
  \citenamefont {Nurujjaman}, \citenamefont {Prakash},\ and\ \citenamefont
  {Prasad~Bal}}]{mahata2021characteristics}%
  \BibitemOpen
  \bibfield  {author} {\bibinfo {author} {\bibfnamefont {A.}~\bibnamefont
  {Mahata}}, \bibinfo {author} {\bibfnamefont {A.}~\bibnamefont {Rai}},
  \bibinfo {author} {\bibfnamefont {M.}~\bibnamefont {Nurujjaman}}, \bibinfo
  {author} {\bibfnamefont {O.}~\bibnamefont {Prakash}}, \ and\ \bibinfo
  {author} {\bibfnamefont {D.}~\bibnamefont {Prasad~Bal}},\ }\bibfield  {title}
  {\enquote {\bibinfo {title} {Characteristics of 2020 stock market crash: The
  covid-19 induced extreme event},}\ }\href@noop {} {\bibfield  {journal}
  {\bibinfo  {journal} {Chaos: An Interdisciplinary Journal of Nonlinear
  Science}\ }\textbf {\bibinfo {volume} {31}},\ \bibinfo {pages} {053115}
  (\bibinfo {year} {2021}{\natexlab{a}})}\BibitemShut {NoStop}%
\bibitem [{\citenamefont {Abe}\ and\ \citenamefont
  {Suzuki}(2004)}]{abe2004aging}%
  \BibitemOpen
  \bibfield  {author} {\bibinfo {author} {\bibfnamefont {S.}~\bibnamefont
  {Abe}}\ and\ \bibinfo {author} {\bibfnamefont {N.}~\bibnamefont {Suzuki}},\
  }\bibfield  {title} {\enquote {\bibinfo {title} {Aging and scaling of
  earthquake aftershocks},}\ }\href@noop {} {\bibfield  {journal} {\bibinfo
  {journal} {Physica A: Statistical Mechanics and its Applications}\ }\textbf
  {\bibinfo {volume} {332}},\ \bibinfo {pages} {533--538} (\bibinfo {year}
  {2004})}\BibitemShut {NoStop}%
\bibitem [{\citenamefont {Siokis}(2012)}]{siokis2012stock}%
  \BibitemOpen
  \bibfield  {author} {\bibinfo {author} {\bibfnamefont {F.~M.}\ \bibnamefont
  {Siokis}},\ }\bibfield  {title} {\enquote {\bibinfo {title} {Stock market
  dynamics: Before and after stock market crashes},}\ }\href@noop {} {\bibfield
   {journal} {\bibinfo  {journal} {Physica A: Statistical Mechanics and its
  Applications}\ }\textbf {\bibinfo {volume} {391}},\ \bibinfo {pages}
  {1315--1322} (\bibinfo {year} {2012})}\BibitemShut {NoStop}%
\bibitem [{\citenamefont {Bak}\ \emph {et~al.}(2002)\citenamefont {Bak},
  \citenamefont {Christensen}, \citenamefont {Danon},\ and\ \citenamefont
  {Scanlon}}]{bak2002unified}%
  \BibitemOpen
  \bibfield  {author} {\bibinfo {author} {\bibfnamefont {P.}~\bibnamefont
  {Bak}}, \bibinfo {author} {\bibfnamefont {K.}~\bibnamefont {Christensen}},
  \bibinfo {author} {\bibfnamefont {L.}~\bibnamefont {Danon}}, \ and\ \bibinfo
  {author} {\bibfnamefont {T.}~\bibnamefont {Scanlon}},\ }\bibfield  {title}
  {\enquote {\bibinfo {title} {Unified scaling law for earthquakes},}\
  }\href@noop {} {\bibfield  {journal} {\bibinfo  {journal} {Physical Review
  Letters}\ }\textbf {\bibinfo {volume} {88}},\ \bibinfo {pages} {178501}
  (\bibinfo {year} {2002})}\BibitemShut {NoStop}%
\bibitem [{\citenamefont {Bai}\ and\ \citenamefont {Zhu}(2010)}]{bai2010power}%
  \BibitemOpen
  \bibfield  {author} {\bibinfo {author} {\bibfnamefont {M.-Y.}\ \bibnamefont
  {Bai}}\ and\ \bibinfo {author} {\bibfnamefont {H.-B.}\ \bibnamefont {Zhu}},\
  }\bibfield  {title} {\enquote {\bibinfo {title} {Power law and multiscaling
  properties of the chinese stock market},}\ }\href@noop {} {\bibfield
  {journal} {\bibinfo  {journal} {Physica A: Statistical Mechanics and its
  Applications}\ }\textbf {\bibinfo {volume} {389}},\ \bibinfo {pages}
  {1883--1890} (\bibinfo {year} {2010})}\BibitemShut {NoStop}%
\bibitem [{\citenamefont {Gabaix}\ \emph {et~al.}(2007)\citenamefont {Gabaix},
  \citenamefont {Gopikrishnan}, \citenamefont {Plerou},\ and\ \citenamefont
  {Stanley}}]{gabaix2007unified}%
  \BibitemOpen
  \bibfield  {author} {\bibinfo {author} {\bibfnamefont {X.}~\bibnamefont
  {Gabaix}}, \bibinfo {author} {\bibfnamefont {P.}~\bibnamefont
  {Gopikrishnan}}, \bibinfo {author} {\bibfnamefont {V.}~\bibnamefont
  {Plerou}}, \ and\ \bibinfo {author} {\bibfnamefont {E.}~\bibnamefont
  {Stanley}},\ }\bibfield  {title} {\enquote {\bibinfo {title} {A unified
  econophysics explanation for the power-law exponents of stock market
  activity},}\ }\href@noop {} {\bibfield  {journal} {\bibinfo  {journal}
  {Physica A: Statistical Mechanics and its Applications}\ }\textbf {\bibinfo
  {volume} {382}},\ \bibinfo {pages} {81--88} (\bibinfo {year}
  {2007})}\BibitemShut {NoStop}%
\bibitem [{\citenamefont {Sel{\c{c}}uk}(2004)}]{selccuk2004financial}%
  \BibitemOpen
  \bibfield  {author} {\bibinfo {author} {\bibfnamefont {F.}~\bibnamefont
  {Sel{\c{c}}uk}},\ }\bibfield  {title} {\enquote {\bibinfo {title} {Financial
  earthquakes, aftershocks and scaling in emerging stock markets},}\
  }\href@noop {} {\bibfield  {journal} {\bibinfo  {journal} {Physica A:
  Statistical Mechanics and its Applications}\ }\textbf {\bibinfo {volume}
  {333}},\ \bibinfo {pages} {306--316} (\bibinfo {year} {2004})}\BibitemShut
  {NoStop}%
\bibitem [{\citenamefont {Kapopoulos}\ and\ \citenamefont
  {Siokis}(2005)}]{kapopoulos2005stock}%
  \BibitemOpen
  \bibfield  {author} {\bibinfo {author} {\bibfnamefont {P.}~\bibnamefont
  {Kapopoulos}}\ and\ \bibinfo {author} {\bibfnamefont {F.}~\bibnamefont
  {Siokis}},\ }\bibfield  {title} {\enquote {\bibinfo {title} {Stock market
  crashes and dynamics of aftershocks},}\ }\href@noop {} {\bibfield  {journal}
  {\bibinfo  {journal} {Economics Letters}\ }\textbf {\bibinfo {volume} {89}},\
  \bibinfo {pages} {48--54} (\bibinfo {year} {2005})}\BibitemShut {NoStop}%
\bibitem [{\citenamefont {Albuquerque}\ \emph {et~al.}(2020)\citenamefont
  {Albuquerque}, \citenamefont {Koskinen}, \citenamefont {Yang},\ and\
  \citenamefont {Zhang}}]{albuquerque2020resiliency}%
  \BibitemOpen
  \bibfield  {author} {\bibinfo {author} {\bibfnamefont {R.}~\bibnamefont
  {Albuquerque}}, \bibinfo {author} {\bibfnamefont {Y.}~\bibnamefont
  {Koskinen}}, \bibinfo {author} {\bibfnamefont {S.}~\bibnamefont {Yang}}, \
  and\ \bibinfo {author} {\bibfnamefont {C.}~\bibnamefont {Zhang}},\ }\bibfield
   {title} {\enquote {\bibinfo {title} {Resiliency of environmental and social
  stocks: an analysis of the exogenous covid-19 market crash},}\ }\href@noop {}
  {\bibfield  {journal} {\bibinfo  {journal} {The Review of Corporate Finance
  Studies}\ }\textbf {\bibinfo {volume} {9}},\ \bibinfo {pages} {593--621}
  (\bibinfo {year} {2020})}\BibitemShut {NoStop}%
\bibitem [{\citenamefont {Mazur}, \citenamefont {Dang},\ and\ \citenamefont
  {Vega}(2020)}]{mazur2020covid}%
  \BibitemOpen
  \bibfield  {author} {\bibinfo {author} {\bibfnamefont {M.}~\bibnamefont
  {Mazur}}, \bibinfo {author} {\bibfnamefont {M.}~\bibnamefont {Dang}}, \ and\
  \bibinfo {author} {\bibfnamefont {M.}~\bibnamefont {Vega}},\ }\bibfield
  {title} {\enquote {\bibinfo {title} {Covid-19 and the march 2020 stock market
  crash. evidence from s\&p1500},}\ }\href@noop {} {\bibfield  {journal}
  {\bibinfo  {journal} {Finance Research Letters}\ ,\ \bibinfo {pages}
  {101690}} (\bibinfo {year} {2020})}\BibitemShut {NoStop}%
\bibitem [{\citenamefont {Ly{\'o}csa}\ \emph {et~al.}(2020)\citenamefont
  {Ly{\'o}csa}, \citenamefont {Baum{\"o}hl}, \citenamefont {V{\`y}rost},\ and\
  \citenamefont {Moln{\'a}r}}]{lyocsa2020fear}%
  \BibitemOpen
  \bibfield  {author} {\bibinfo {author} {\bibfnamefont {{\v{S}}.}~\bibnamefont
  {Ly{\'o}csa}}, \bibinfo {author} {\bibfnamefont {E.}~\bibnamefont
  {Baum{\"o}hl}}, \bibinfo {author} {\bibfnamefont {T.}~\bibnamefont
  {V{\`y}rost}}, \ and\ \bibinfo {author} {\bibfnamefont {P.}~\bibnamefont
  {Moln{\'a}r}},\ }\bibfield  {title} {\enquote {\bibinfo {title} {Fear of the
  coronavirus and the stock markets},}\ }\href@noop {} {\bibfield  {journal}
  {\bibinfo  {journal} {Finance research letters}\ }\textbf {\bibinfo {volume}
  {36}},\ \bibinfo {pages} {101735} (\bibinfo {year} {2020})}\BibitemShut
  {NoStop}%
\bibitem [{\citenamefont {Baker}\ \emph {et~al.}(2020)\citenamefont {Baker},
  \citenamefont {Bloom}, \citenamefont {Davis}, \citenamefont {Kost},
  \citenamefont {Sammon},\ and\ \citenamefont
  {Viratyosin}}]{baker2020unprecedented}%
  \BibitemOpen
  \bibfield  {author} {\bibinfo {author} {\bibfnamefont {S.~R.}\ \bibnamefont
  {Baker}}, \bibinfo {author} {\bibfnamefont {N.}~\bibnamefont {Bloom}},
  \bibinfo {author} {\bibfnamefont {S.~J.}\ \bibnamefont {Davis}}, \bibinfo
  {author} {\bibfnamefont {K.~J.}\ \bibnamefont {Kost}}, \bibinfo {author}
  {\bibfnamefont {M.~C.}\ \bibnamefont {Sammon}}, \ and\ \bibinfo {author}
  {\bibfnamefont {T.}~\bibnamefont {Viratyosin}},\ }\href@noop {} {\enquote
  {\bibinfo {title} {The unprecedented stock market impact of covid-19},}\
  }\bibinfo {type} {Tech. Rep.}\ (\bibinfo  {institution} {National Bureau of
  Economic Research},\ \bibinfo {year} {2020})\BibitemShut {NoStop}%
\bibitem [{\citenamefont {Wang}\ \emph {et~al.}(2009)\citenamefont {Wang},
  \citenamefont {Meric}, \citenamefont {Liu},\ and\ \citenamefont
  {Meric}}]{wang2009stock}%
  \BibitemOpen
  \bibfield  {author} {\bibinfo {author} {\bibfnamefont {J.}~\bibnamefont
  {Wang}}, \bibinfo {author} {\bibfnamefont {G.}~\bibnamefont {Meric}},
  \bibinfo {author} {\bibfnamefont {Z.}~\bibnamefont {Liu}}, \ and\ \bibinfo
  {author} {\bibfnamefont {I.}~\bibnamefont {Meric}},\ }\bibfield  {title}
  {\enquote {\bibinfo {title} {Stock market crashes, firm characteristics, and
  stock returns},}\ }\href@noop {} {\bibfield  {journal} {\bibinfo  {journal}
  {Journal of Banking \& Finance}\ }\textbf {\bibinfo {volume} {33}},\ \bibinfo
  {pages} {1563--1574} (\bibinfo {year} {2009})}\BibitemShut {NoStop}%
\bibitem [{\citenamefont {Barlevy}\ and\ \citenamefont
  {Veronesi}(2003)}]{barlevy2003rational}%
  \BibitemOpen
  \bibfield  {author} {\bibinfo {author} {\bibfnamefont {G.}~\bibnamefont
  {Barlevy}}\ and\ \bibinfo {author} {\bibfnamefont {P.}~\bibnamefont
  {Veronesi}},\ }\bibfield  {title} {\enquote {\bibinfo {title} {Rational
  panics and stock market crashes},}\ }\href@noop {} {\bibfield  {journal}
  {\bibinfo  {journal} {Journal of Economic Theory}\ }\textbf {\bibinfo
  {volume} {110}},\ \bibinfo {pages} {234--263} (\bibinfo {year}
  {2003})}\BibitemShut {NoStop}%
\bibitem [{\citenamefont {Chiang}\ and\ \citenamefont
  {Zheng}(2010)}]{chiang2010empirical}%
  \BibitemOpen
  \bibfield  {author} {\bibinfo {author} {\bibfnamefont {T.~C.}\ \bibnamefont
  {Chiang}}\ and\ \bibinfo {author} {\bibfnamefont {D.}~\bibnamefont {Zheng}},\
  }\bibfield  {title} {\enquote {\bibinfo {title} {An empirical analysis of
  herd behavior in global stock markets},}\ }\href@noop {} {\bibfield
  {journal} {\bibinfo  {journal} {Journal of Banking \& Finance}\ }\textbf
  {\bibinfo {volume} {34}},\ \bibinfo {pages} {1911--1921} (\bibinfo {year}
  {2010})}\BibitemShut {NoStop}%
\bibitem [{\citenamefont {Tan}\ \emph {et~al.}(2008)\citenamefont {Tan},
  \citenamefont {Chiang}, \citenamefont {Mason},\ and\ \citenamefont
  {Nelling}}]{tan2008herding}%
  \BibitemOpen
  \bibfield  {author} {\bibinfo {author} {\bibfnamefont {L.}~\bibnamefont
  {Tan}}, \bibinfo {author} {\bibfnamefont {T.~C.}\ \bibnamefont {Chiang}},
  \bibinfo {author} {\bibfnamefont {J.~R.}\ \bibnamefont {Mason}}, \ and\
  \bibinfo {author} {\bibfnamefont {E.}~\bibnamefont {Nelling}},\ }\bibfield
  {title} {\enquote {\bibinfo {title} {Herding behavior in chinese stock
  markets: An examination of a and b shares},}\ }\href@noop {} {\bibfield
  {journal} {\bibinfo  {journal} {Pacific-Basin Finance Journal}\ }\textbf
  {\bibinfo {volume} {16}},\ \bibinfo {pages} {61--77} (\bibinfo {year}
  {2008})}\BibitemShut {NoStop}%
\bibitem [{\citenamefont {White}(1990)}]{white1990stock}%
  \BibitemOpen
  \bibfield  {author} {\bibinfo {author} {\bibfnamefont {E.~N.}\ \bibnamefont
  {White}},\ }\bibfield  {title} {\enquote {\bibinfo {title} {The stock market
  boom and crash of 1929 revisited},}\ }\href@noop {} {\bibfield  {journal}
  {\bibinfo  {journal} {Journal of Economic perspectives}\ }\textbf {\bibinfo
  {volume} {4}},\ \bibinfo {pages} {67--83} (\bibinfo {year}
  {1990})}\BibitemShut {NoStop}%
\bibitem [{\citenamefont {Choudhry}(1996)}]{choudhry1996stock}%
  \BibitemOpen
  \bibfield  {author} {\bibinfo {author} {\bibfnamefont {T.}~\bibnamefont
  {Choudhry}},\ }\bibfield  {title} {\enquote {\bibinfo {title} {Stock market
  volatility and the crash of 1987: evidence from six emerging markets},}\
  }\href@noop {} {\bibfield  {journal} {\bibinfo  {journal} {Journal of
  International money and Finance}\ }\textbf {\bibinfo {volume} {15}},\
  \bibinfo {pages} {969--981} (\bibinfo {year} {1996})}\BibitemShut {NoStop}%
\bibitem [{\citenamefont {Carlson}(2007)}]{carlson2007brief}%
  \BibitemOpen
  \bibfield  {author} {\bibinfo {author} {\bibfnamefont {M.~A.}\ \bibnamefont
  {Carlson}},\ }\bibfield  {title} {\enquote {\bibinfo {title} {A brief history
  of the 1987 stock market crash with a discussion of the federal reserve
  response},}\ }\href@noop {} {\  (\bibinfo {year} {2007})}\BibitemShut
  {NoStop}%
\bibitem [{\citenamefont {Grant}(1990)}]{grant1990stock}%
  \BibitemOpen
  \bibfield  {author} {\bibinfo {author} {\bibfnamefont {J.~L.}\ \bibnamefont
  {Grant}},\ }\bibfield  {title} {\enquote {\bibinfo {title} {Stock return
  volatility during the crash of 1987},}\ }\href@noop {} {\bibfield  {journal}
  {\bibinfo  {journal} {Journal of Portfolio Management}\ }\textbf {\bibinfo
  {volume} {16}},\ \bibinfo {pages} {69} (\bibinfo {year} {1990})}\BibitemShut
  {NoStop}%
\bibitem [{\citenamefont {Sornette}\ and\ \citenamefont
  {Johansen}(1997)}]{sornette1997large}%
  \BibitemOpen
  \bibfield  {author} {\bibinfo {author} {\bibfnamefont {D.}~\bibnamefont
  {Sornette}}\ and\ \bibinfo {author} {\bibfnamefont {A.}~\bibnamefont
  {Johansen}},\ }\bibfield  {title} {\enquote {\bibinfo {title} {Large
  financial crashes},}\ }\href@noop {} {\bibfield  {journal} {\bibinfo
  {journal} {Physica A: Statistical Mechanics and its Applications}\ }\textbf
  {\bibinfo {volume} {245}},\ \bibinfo {pages} {411--422} (\bibinfo {year}
  {1997})}\BibitemShut {NoStop}%
\bibitem [{\citenamefont {Mahata}\ \emph
  {et~al.}(2021{\natexlab{b}})\citenamefont {Mahata}, \citenamefont {Rai},
  \citenamefont {Nurujjaman},\ and\ \citenamefont
  {Prakash}}]{mahata2021modeling}%
  \BibitemOpen
  \bibfield  {author} {\bibinfo {author} {\bibfnamefont {A.}~\bibnamefont
  {Mahata}}, \bibinfo {author} {\bibfnamefont {A.}~\bibnamefont {Rai}},
  \bibinfo {author} {\bibfnamefont {M.}~\bibnamefont {Nurujjaman}}, \ and\
  \bibinfo {author} {\bibfnamefont {O.}~\bibnamefont {Prakash}},\ }\bibfield
  {title} {\enquote {\bibinfo {title} {Modeling and analysis of the effect of
  covid-19 on the stock price: V and l-shape recovery},}\ }\href@noop {}
  {\bibfield  {journal} {\bibinfo  {journal} {Physica A: Statistical Mechanics
  and its Applications}\ }\textbf {\bibinfo {volume} {574}},\ \bibinfo {pages}
  {126008} (\bibinfo {year} {2021}{\natexlab{b}})}\BibitemShut {NoStop}%
\bibitem [{\citenamefont {Netter}\ and\ \citenamefont
  {Mitchell}(1989)}]{netter1989stock}%
  \BibitemOpen
  \bibfield  {author} {\bibinfo {author} {\bibfnamefont {J.~M.}\ \bibnamefont
  {Netter}}\ and\ \bibinfo {author} {\bibfnamefont {M.~L.}\ \bibnamefont
  {Mitchell}},\ }\bibfield  {title} {\enquote {\bibinfo {title}
  {Stock-repurchase announcements and insider transactions after the october
  1987 stock market crash},}\ }\href@noop {} {\bibfield  {journal} {\bibinfo
  {journal} {Financial Management}\ ,\ \bibinfo {pages} {84--96}} (\bibinfo
  {year} {1989})}\BibitemShut {NoStop}%
\bibitem [{\citenamefont {Hong}\ and\ \citenamefont
  {Stein}(2003)}]{hong2003differences}%
  \BibitemOpen
  \bibfield  {author} {\bibinfo {author} {\bibfnamefont {H.}~\bibnamefont
  {Hong}}\ and\ \bibinfo {author} {\bibfnamefont {J.~C.}\ \bibnamefont
  {Stein}},\ }\bibfield  {title} {\enquote {\bibinfo {title} {Differences of
  opinion, short-sales constraints, and market crashes},}\ }\href@noop {}
  {\bibfield  {journal} {\bibinfo  {journal} {The Review of Financial Studies}\
  }\textbf {\bibinfo {volume} {16}},\ \bibinfo {pages} {487--525} (\bibinfo
  {year} {2003})}\BibitemShut {NoStop}%
\bibitem [{\citenamefont {Gopikrishnan}\ \emph {et~al.}(2000)\citenamefont
  {Gopikrishnan}, \citenamefont {Plerou}, \citenamefont {Gabaix},\ and\
  \citenamefont {Stanley}}]{gopikrishnan2000statistical}%
  \BibitemOpen
  \bibfield  {author} {\bibinfo {author} {\bibfnamefont {P.}~\bibnamefont
  {Gopikrishnan}}, \bibinfo {author} {\bibfnamefont {V.}~\bibnamefont
  {Plerou}}, \bibinfo {author} {\bibfnamefont {X.}~\bibnamefont {Gabaix}}, \
  and\ \bibinfo {author} {\bibfnamefont {H.~E.}\ \bibnamefont {Stanley}},\
  }\bibfield  {title} {\enquote {\bibinfo {title} {Statistical properties of
  share volume traded in financial markets},}\ }\href@noop {} {\bibfield
  {journal} {\bibinfo  {journal} {Physical review e}\ }\textbf {\bibinfo
  {volume} {62}},\ \bibinfo {pages} {R4493} (\bibinfo {year}
  {2000})}\BibitemShut {NoStop}%
\bibitem [{\citenamefont {Oh}\ \emph {et~al.}(2011)\citenamefont {Oh},
  \citenamefont {Eom}, \citenamefont {Wang}, \citenamefont {Jung},
  \citenamefont {Stanley},\ and\ \citenamefont {Kim}}]{oh2011statistical}%
  \BibitemOpen
  \bibfield  {author} {\bibinfo {author} {\bibfnamefont {G.}~\bibnamefont
  {Oh}}, \bibinfo {author} {\bibfnamefont {C.}~\bibnamefont {Eom}}, \bibinfo
  {author} {\bibfnamefont {F.}~\bibnamefont {Wang}}, \bibinfo {author}
  {\bibfnamefont {W.-S.}\ \bibnamefont {Jung}}, \bibinfo {author}
  {\bibfnamefont {H.~E.}\ \bibnamefont {Stanley}}, \ and\ \bibinfo {author}
  {\bibfnamefont {S.}~\bibnamefont {Kim}},\ }\bibfield  {title} {\enquote
  {\bibinfo {title} {Statistical properties of cross-correlation in the korean
  stock market},}\ }\href@noop {} {\bibfield  {journal} {\bibinfo  {journal}
  {The European Physical Journal B}\ }\textbf {\bibinfo {volume} {79}},\
  \bibinfo {pages} {55--60} (\bibinfo {year} {2011})}\BibitemShut {NoStop}%
\bibitem [{\citenamefont {Liu}\ \emph {et~al.}(1999)\citenamefont {Liu},
  \citenamefont {Gopikrishnan}, \citenamefont {Stanley} \emph
  {et~al.}}]{liu1999statistical}%
  \BibitemOpen
  \bibfield  {author} {\bibinfo {author} {\bibfnamefont {Y.}~\bibnamefont
  {Liu}}, \bibinfo {author} {\bibfnamefont {P.}~\bibnamefont {Gopikrishnan}},
  \bibinfo {author} {\bibfnamefont {H.~E.}\ \bibnamefont {Stanley}},  \emph
  {et~al.},\ }\bibfield  {title} {\enquote {\bibinfo {title} {Statistical
  properties of the volatility of price fluctuations},}\ }\href@noop {}
  {\bibfield  {journal} {\bibinfo  {journal} {Physical review e}\ }\textbf
  {\bibinfo {volume} {60}},\ \bibinfo {pages} {1390} (\bibinfo {year}
  {1999})}\BibitemShut {NoStop}%
\bibitem [{\citenamefont {Mantegna}\ and\ \citenamefont
  {Stanley}(1997)}]{mantegna1997stock}%
  \BibitemOpen
  \bibfield  {author} {\bibinfo {author} {\bibfnamefont {R.~N.}\ \bibnamefont
  {Mantegna}}\ and\ \bibinfo {author} {\bibfnamefont {H.~E.}\ \bibnamefont
  {Stanley}},\ }\bibfield  {title} {\enquote {\bibinfo {title} {Stock market
  dynamics and turbulence: parallel analysis of fluctuation phenomena},}\
  }\href@noop {} {\bibfield  {journal} {\bibinfo  {journal} {Physica A:
  Statistical Mechanics and its Applications}\ }\textbf {\bibinfo {volume}
  {239}},\ \bibinfo {pages} {255--266} (\bibinfo {year} {1997})}\BibitemShut
  {NoStop}%
\bibitem [{\citenamefont {Cont}(2001)}]{cont2001empirical}%
  \BibitemOpen
  \bibfield  {author} {\bibinfo {author} {\bibfnamefont {R.}~\bibnamefont
  {Cont}},\ }\bibfield  {title} {\enquote {\bibinfo {title} {Empirical
  properties of asset returns: stylized facts and statistical issues},}\
  }\href@noop {} {\  (\bibinfo {year} {2001})}\BibitemShut {NoStop}%
\bibitem [{\citenamefont {Restocchi}, \citenamefont {McGroarty},\ and\
  \citenamefont {Gerding}(2019)}]{restocchi2019statistical}%
  \BibitemOpen
  \bibfield  {author} {\bibinfo {author} {\bibfnamefont {V.}~\bibnamefont
  {Restocchi}}, \bibinfo {author} {\bibfnamefont {F.}~\bibnamefont
  {McGroarty}}, \ and\ \bibinfo {author} {\bibfnamefont {E.}~\bibnamefont
  {Gerding}},\ }\bibfield  {title} {\enquote {\bibinfo {title} {Statistical
  properties of volume and calendar effects in prediction markets},}\
  }\href@noop {} {\bibfield  {journal} {\bibinfo  {journal} {Physica A:
  Statistical Mechanics and its Applications}\ }\textbf {\bibinfo {volume}
  {523}},\ \bibinfo {pages} {1150--1160} (\bibinfo {year} {2019})}\BibitemShut
  {NoStop}%
\bibitem [{\citenamefont {Da~Cunha}\ and\ \citenamefont
  {da~Silva}(2020)}]{da2020relevant}%
  \BibitemOpen
  \bibfield  {author} {\bibinfo {author} {\bibfnamefont {C.}~\bibnamefont
  {Da~Cunha}}\ and\ \bibinfo {author} {\bibfnamefont {R.}~\bibnamefont
  {da~Silva}},\ }\bibfield  {title} {\enquote {\bibinfo {title} {Relevant
  stylized facts about bitcoin: Fluctuations, first return probability, and
  natural phenomena},}\ }\href@noop {} {\bibfield  {journal} {\bibinfo
  {journal} {Physica A: Statistical Mechanics and its Applications}\ }\textbf
  {\bibinfo {volume} {550}},\ \bibinfo {pages} {124155} (\bibinfo {year}
  {2020})}\BibitemShut {NoStop}%
\bibitem [{\citenamefont {Noh}(2000)}]{noh2000model}%
  \BibitemOpen
  \bibfield  {author} {\bibinfo {author} {\bibfnamefont {J.~D.}\ \bibnamefont
  {Noh}},\ }\bibfield  {title} {\enquote {\bibinfo {title} {Model for
  correlations in stock markets},}\ }\href@noop {} {\bibfield  {journal}
  {\bibinfo  {journal} {Physical Review E}\ }\textbf {\bibinfo {volume} {61}},\
  \bibinfo {pages} {5981} (\bibinfo {year} {2000})}\BibitemShut {NoStop}%
\bibitem [{\citenamefont {Hamao}, \citenamefont {Masulis},\ and\ \citenamefont
  {Ng}(1990)}]{hamao1990correlations}%
  \BibitemOpen
  \bibfield  {author} {\bibinfo {author} {\bibfnamefont {Y.}~\bibnamefont
  {Hamao}}, \bibinfo {author} {\bibfnamefont {R.~W.}\ \bibnamefont {Masulis}},
  \ and\ \bibinfo {author} {\bibfnamefont {V.}~\bibnamefont {Ng}},\ }\bibfield
  {title} {\enquote {\bibinfo {title} {Correlations in price changes and
  volatility across international stock markets},}\ }\href@noop {} {\bibfield
  {journal} {\bibinfo  {journal} {The review of financial studies}\ }\textbf
  {\bibinfo {volume} {3}},\ \bibinfo {pages} {281--307} (\bibinfo {year}
  {1990})}\BibitemShut {NoStop}%
\bibitem [{\citenamefont {Wang}, \citenamefont {Shang},\ and\ \citenamefont
  {Ge}(2012)}]{wang2012multifractal}%
  \BibitemOpen
  \bibfield  {author} {\bibinfo {author} {\bibfnamefont {J.}~\bibnamefont
  {Wang}}, \bibinfo {author} {\bibfnamefont {P.}~\bibnamefont {Shang}}, \ and\
  \bibinfo {author} {\bibfnamefont {W.}~\bibnamefont {Ge}},\ }\bibfield
  {title} {\enquote {\bibinfo {title} {Multifractal cross-correlation analysis
  based on statistical moments},}\ }\href@noop {} {\bibfield  {journal}
  {\bibinfo  {journal} {Fractals}\ }\textbf {\bibinfo {volume} {20}},\ \bibinfo
  {pages} {271--279} (\bibinfo {year} {2012})}\BibitemShut {NoStop}%
\bibitem [{\citenamefont {Turiel}\ and\ \citenamefont
  {P{\'e}rez-Vicente}(2003)}]{turiel2003multifractal}%
  \BibitemOpen
  \bibfield  {author} {\bibinfo {author} {\bibfnamefont {A.}~\bibnamefont
  {Turiel}}\ and\ \bibinfo {author} {\bibfnamefont {C.~J.}\ \bibnamefont
  {P{\'e}rez-Vicente}},\ }\bibfield  {title} {\enquote {\bibinfo {title}
  {Multifractal geometry in stock market time series},}\ }\href@noop {}
  {\bibfield  {journal} {\bibinfo  {journal} {Physica A: Statistical Mechanics
  and its Applications}\ }\textbf {\bibinfo {volume} {322}},\ \bibinfo {pages}
  {629--649} (\bibinfo {year} {2003})}\BibitemShut {NoStop}%
\bibitem [{\citenamefont {Potirakis}, \citenamefont {Zitis},\ and\
  \citenamefont {Eftaxias}(2013)}]{potirakis2013dynamical}%
  \BibitemOpen
  \bibfield  {author} {\bibinfo {author} {\bibfnamefont {S.~M.}\ \bibnamefont
  {Potirakis}}, \bibinfo {author} {\bibfnamefont {P.~I.}\ \bibnamefont
  {Zitis}}, \ and\ \bibinfo {author} {\bibfnamefont {K.}~\bibnamefont
  {Eftaxias}},\ }\bibfield  {title} {\enquote {\bibinfo {title} {Dynamical
  analogy between economical crisis and earthquake dynamics within the
  nonextensive statistical mechanics framework},}\ }\href@noop {} {\bibfield
  {journal} {\bibinfo  {journal} {Physica A: Statistical Mechanics and its
  Applications}\ }\textbf {\bibinfo {volume} {392}},\ \bibinfo {pages}
  {2940--2954} (\bibinfo {year} {2013})}\BibitemShut {NoStop}%
\bibitem [{\citenamefont {Petersen}\ \emph {et~al.}(2010)\citenamefont
  {Petersen}, \citenamefont {Wang}, \citenamefont {Havlin},\ and\ \citenamefont
  {Stanley}}]{petersen2010market}%
  \BibitemOpen
  \bibfield  {author} {\bibinfo {author} {\bibfnamefont {A.~M.}\ \bibnamefont
  {Petersen}}, \bibinfo {author} {\bibfnamefont {F.}~\bibnamefont {Wang}},
  \bibinfo {author} {\bibfnamefont {S.}~\bibnamefont {Havlin}}, \ and\ \bibinfo
  {author} {\bibfnamefont {H.~E.}\ \bibnamefont {Stanley}},\ }\bibfield
  {title} {\enquote {\bibinfo {title} {Market dynamics immediately before and
  after financial shocks: Quantifying the omori, productivity, and bath
  laws},}\ }\href@noop {} {\bibfield  {journal} {\bibinfo  {journal} {Physical
  Review E}\ }\textbf {\bibinfo {volume} {82}},\ \bibinfo {pages} {036114}
  (\bibinfo {year} {2010})}\BibitemShut {NoStop}%
\bibitem [{\citenamefont {Lillo}\ and\ \citenamefont
  {Mantegna}(2003)}]{lillo2003power}%
  \BibitemOpen
  \bibfield  {author} {\bibinfo {author} {\bibfnamefont {F.}~\bibnamefont
  {Lillo}}\ and\ \bibinfo {author} {\bibfnamefont {R.~N.}\ \bibnamefont
  {Mantegna}},\ }\bibfield  {title} {\enquote {\bibinfo {title} {Power-law
  relaxation in a complex system: Omori law after a financial market crash},}\
  }\href@noop {} {\bibfield  {journal} {\bibinfo  {journal} {Physical Review
  E}\ }\textbf {\bibinfo {volume} {68}},\ \bibinfo {pages} {016119} (\bibinfo
  {year} {2003})}\BibitemShut {NoStop}%
\bibitem [{\citenamefont {Mu}\ and\ \citenamefont
  {Zhou}(2008)}]{mu2008relaxation}%
  \BibitemOpen
  \bibfield  {author} {\bibinfo {author} {\bibfnamefont {G.-H.}\ \bibnamefont
  {Mu}}\ and\ \bibinfo {author} {\bibfnamefont {W.-X.}\ \bibnamefont {Zhou}},\
  }\bibfield  {title} {\enquote {\bibinfo {title} {Relaxation dynamics of
  aftershocks after large volatility shocks in the ssec index},}\ }\href@noop
  {} {\bibfield  {journal} {\bibinfo  {journal} {Physica A: Statistical
  Mechanics and its Applications}\ }\textbf {\bibinfo {volume} {387}},\
  \bibinfo {pages} {5211--5218} (\bibinfo {year} {2008})}\BibitemShut {NoStop}%
\bibitem [{\citenamefont {Gresnigt}, \citenamefont {Kole},\ and\ \citenamefont
  {Franses}(2015)}]{gresnigt2015interpreting}%
  \BibitemOpen
  \bibfield  {author} {\bibinfo {author} {\bibfnamefont {F.}~\bibnamefont
  {Gresnigt}}, \bibinfo {author} {\bibfnamefont {E.}~\bibnamefont {Kole}}, \
  and\ \bibinfo {author} {\bibfnamefont {P.~H.}\ \bibnamefont {Franses}},\
  }\bibfield  {title} {\enquote {\bibinfo {title} {Interpreting financial
  market crashes as earthquakes: A new early warning system for medium term
  crashes},}\ }\href@noop {} {\bibfield  {journal} {\bibinfo  {journal}
  {Journal of Banking \& Finance}\ }\textbf {\bibinfo {volume} {56}},\ \bibinfo
  {pages} {123--139} (\bibinfo {year} {2015})}\BibitemShut {NoStop}%
\bibitem [{\citenamefont {Bai}\ and\ \citenamefont
  {Perron}(1998)}]{bai1998estimating}%
  \BibitemOpen
  \bibfield  {author} {\bibinfo {author} {\bibfnamefont {J.}~\bibnamefont
  {Bai}}\ and\ \bibinfo {author} {\bibfnamefont {P.}~\bibnamefont {Perron}},\
  }\bibfield  {title} {\enquote {\bibinfo {title} {Estimating and testing
  linear models with multiple structural changes},}\ }\href@noop {} {\bibfield
  {journal} {\bibinfo  {journal} {Econometrica}\ ,\ \bibinfo {pages} {47--78}}
  (\bibinfo {year} {1998})}\BibitemShut {NoStop}%
\bibitem [{\citenamefont {Bai}\ and\ \citenamefont
  {Perron}(2003)}]{bai2003computation}%
  \BibitemOpen
  \bibfield  {author} {\bibinfo {author} {\bibfnamefont {J.}~\bibnamefont
  {Bai}}\ and\ \bibinfo {author} {\bibfnamefont {P.}~\bibnamefont {Perron}},\
  }\bibfield  {title} {\enquote {\bibinfo {title} {Computation and analysis of
  multiple structural change models},}\ }\href@noop {} {\bibfield  {journal}
  {\bibinfo  {journal} {Journal of applied econometrics}\ }\textbf {\bibinfo
  {volume} {18}},\ \bibinfo {pages} {1--22} (\bibinfo {year}
  {2003})}\BibitemShut {NoStop}%
\bibitem [{\citenamefont {Clauset}, \citenamefont {Shalizi},\ and\
  \citenamefont {Newman}(2009)}]{clauset2009power}%
  \BibitemOpen
  \bibfield  {author} {\bibinfo {author} {\bibfnamefont {A.}~\bibnamefont
  {Clauset}}, \bibinfo {author} {\bibfnamefont {C.~R.}\ \bibnamefont
  {Shalizi}}, \ and\ \bibinfo {author} {\bibfnamefont {M.~E.}\ \bibnamefont
  {Newman}},\ }\bibfield  {title} {\enquote {\bibinfo {title} {Power-law
  distributions in empirical data},}\ }\href@noop {} {\bibfield  {journal}
  {\bibinfo  {journal} {SIAM review}\ }\textbf {\bibinfo {volume} {51}},\
  \bibinfo {pages} {661--703} (\bibinfo {year} {2009})}\BibitemShut {NoStop}%
\bibitem [{\citenamefont {Taleghani}, \citenamefont {Salehi},\ and\
  \citenamefont {Shakibaiee}(2019)}]{taleghani2019analysis}%
  \BibitemOpen
  \bibfield  {author} {\bibinfo {author} {\bibfnamefont {F.}~\bibnamefont
  {Taleghani}}, \bibinfo {author} {\bibfnamefont {M.}~\bibnamefont {Salehi}}, \
  and\ \bibinfo {author} {\bibfnamefont {A.}~\bibnamefont {Shakibaiee}},\
  }\bibfield  {title} {\enquote {\bibinfo {title} {An analysis of the repeated
  financial earthquakes},}\ }\href@noop {} {\bibfield  {journal} {\bibinfo
  {journal} {Advances in Mathematical Finance and Applications}\ }\textbf
  {\bibinfo {volume} {4}},\ \bibinfo {pages} {59--76} (\bibinfo {year}
  {2019})}\BibitemShut {NoStop}%
\bibitem [{\citenamefont {Sornette}\ and\ \citenamefont
  {Sornette}(1999)}]{sornette1999general}%
  \BibitemOpen
  \bibfield  {author} {\bibinfo {author} {\bibfnamefont {D.}~\bibnamefont
  {Sornette}}\ and\ \bibinfo {author} {\bibfnamefont {A.}~\bibnamefont
  {Sornette}},\ }\bibfield  {title} {\enquote {\bibinfo {title} {General theory
  of the modified gutenberg-richter law for large seismic moments},}\
  }\href@noop {} {\bibfield  {journal} {\bibinfo  {journal} {Bulletin of the
  Seismological Society of America}\ }\textbf {\bibinfo {volume} {89}},\
  \bibinfo {pages} {1121--1130} (\bibinfo {year} {1999})}\BibitemShut {NoStop}%
\bibitem [{\citenamefont {Gutenberg}\ and\ \citenamefont
  {Richter}(1944)}]{gutenberg1944frequency}%
  \BibitemOpen
  \bibfield  {author} {\bibinfo {author} {\bibfnamefont {B.}~\bibnamefont
  {Gutenberg}}\ and\ \bibinfo {author} {\bibfnamefont {C.~F.}\ \bibnamefont
  {Richter}},\ }\bibfield  {title} {\enquote {\bibinfo {title} {Frequency of
  earthquakes in california},}\ }\href@noop {} {\bibfield  {journal} {\bibinfo
  {journal} {Bulletin of the Seismological Society of America}\ }\textbf
  {\bibinfo {volume} {34}},\ \bibinfo {pages} {185--188} (\bibinfo {year}
  {1944})}\BibitemShut {NoStop}%
\bibitem [{\citenamefont {Chowdhury}\ \emph {et~al.}(2017)\citenamefont
  {Chowdhury}, \citenamefont {Deb}, \citenamefont {Nurujjaman},\ and\
  \citenamefont {Barman}}]{chowdhury2017identification}%
  \BibitemOpen
  \bibfield  {author} {\bibinfo {author} {\bibfnamefont {S.}~\bibnamefont
  {Chowdhury}}, \bibinfo {author} {\bibfnamefont {A.}~\bibnamefont {Deb}},
  \bibinfo {author} {\bibfnamefont {M.}~\bibnamefont {Nurujjaman}}, \ and\
  \bibinfo {author} {\bibfnamefont {C.}~\bibnamefont {Barman}},\ }\bibfield
  {title} {\enquote {\bibinfo {title} {Identification of pre-seismic anomalies
  of soil radon-222 signal using hilbert--huang transform},}\ }\href@noop {}
  {\bibfield  {journal} {\bibinfo  {journal} {Natural Hazards}\ }\textbf
  {\bibinfo {volume} {87}},\ \bibinfo {pages} {1587--1606} (\bibinfo {year}
  {2017})}\BibitemShut {NoStop}%
\bibitem [{\citenamefont {Huang}\ \emph {et~al.}(1998)\citenamefont {Huang},
  \citenamefont {Shen}, \citenamefont {Long}, \citenamefont {Wu}, \citenamefont
  {Shih}, \citenamefont {Zheng}, \citenamefont {Yen}, \citenamefont {Tung},\
  and\ \citenamefont {Liu}}]{huang1998empirical}%
  \BibitemOpen
  \bibfield  {author} {\bibinfo {author} {\bibfnamefont {N.~E.}\ \bibnamefont
  {Huang}}, \bibinfo {author} {\bibfnamefont {Z.}~\bibnamefont {Shen}},
  \bibinfo {author} {\bibfnamefont {S.~R.}\ \bibnamefont {Long}}, \bibinfo
  {author} {\bibfnamefont {M.~C.}\ \bibnamefont {Wu}}, \bibinfo {author}
  {\bibfnamefont {H.~H.}\ \bibnamefont {Shih}}, \bibinfo {author}
  {\bibfnamefont {Q.}~\bibnamefont {Zheng}}, \bibinfo {author} {\bibfnamefont
  {N.-C.}\ \bibnamefont {Yen}}, \bibinfo {author} {\bibfnamefont {C.~C.}\
  \bibnamefont {Tung}}, \ and\ \bibinfo {author} {\bibfnamefont {H.~H.}\
  \bibnamefont {Liu}},\ }\bibfield  {title} {\enquote {\bibinfo {title} {The
  empirical mode decomposition and the hilbert spectrum for nonlinear and
  non-stationary time series analysis},}\ }\href@noop {} {\bibfield  {journal}
  {\bibinfo  {journal} {Proceedings of the Royal Society of London. Series A:
  mathematical, physical and engineering sciences}\ }\textbf {\bibinfo {volume}
  {454}},\ \bibinfo {pages} {903--995} (\bibinfo {year} {1998})}\BibitemShut
  {NoStop}%
\bibitem [{\citenamefont {Mahata}, \citenamefont {Bal},\ and\ \citenamefont
  {Nurujjaman}(2020)}]{MAHATA2020123612}%
  \BibitemOpen
  \bibfield  {author} {\bibinfo {author} {\bibfnamefont {A.}~\bibnamefont
  {Mahata}}, \bibinfo {author} {\bibfnamefont {D.~P.}\ \bibnamefont {Bal}}, \
  and\ \bibinfo {author} {\bibfnamefont {M.}~\bibnamefont {Nurujjaman}},\
  }\bibfield  {title} {\enquote {\bibinfo {title} {Identification of short-term
  and long-term time scales in stock markets and effect of structural break},}\
  }\href {\doibase https://doi.org/10.1016/j.physa.2019.123612} {\bibfield
  {journal} {\bibinfo  {journal} {Physica A: Statistical Mechanics and its
  Applications}\ }\textbf {\bibinfo {volume} {545}},\ \bibinfo {pages} {123612}
  (\bibinfo {year} {2020})}\BibitemShut {NoStop}%
\bibitem [{Yah()}]{Yahoofi}%
  \BibitemOpen
  \href@noop {} {}\bibinfo {howpublished}
  {\url{https://in.finance.yahoo.com/}}\BibitemShut {NoStop}%
\bibitem [{\citenamefont {Reiss}, \citenamefont {Thomas},\ and\ \citenamefont
  {Reiss}(1997)}]{reiss1997statistical}%
  \BibitemOpen
  \bibfield  {author} {\bibinfo {author} {\bibfnamefont {R.-D.}\ \bibnamefont
  {Reiss}}, \bibinfo {author} {\bibfnamefont {M.}~\bibnamefont {Thomas}}, \
  and\ \bibinfo {author} {\bibfnamefont {R.}~\bibnamefont {Reiss}},\
  }\href@noop {} {\emph {\bibinfo {title} {Statistical analysis of extreme
  values}}},\ Vol.~\bibinfo {volume} {2}\ (\bibinfo  {publisher} {Springer},\
  \bibinfo {year} {1997})\BibitemShut {NoStop}%
\end{thebibliography}%
\end{document}